\documentclass[%
showpacs,
 amsmath,amssymb,
 aps,
 pra,
floatfix,
]{revtex4-1}

\usepackage{graphics}
\usepackage{dcolumn}
\usepackage{bm}

\usepackage{tikz}
\usepackage[percent]{overpic}

\newcommand{\la}{\langle}
\newcommand{\ra}{\rangle}
\newcommand{\br}{\langle}
\newcommand{\ke}{\rangle}

\newcommand{\beq}{\begin{equation}}
\newcommand{\eeq}{\end{equation}}
\newcommand{\bea}{\begin{eqnarray}}
\newcommand{\eea}{\end{eqnarray}}
\newcommand{\beqa}{\begin{eqnarray}}
\newcommand{\eeqa}{\end{eqnarray}}
\newcommand{\n}{\nonumber \\}

\newcommand\ion[2]{#1$\;${\scshape{#2}}}

\begin{document}

\preprint{APS/123-QED}

\title{Excitation and charge transfer in low-energy hydrogen atom collisions with neutral atoms:
Theory, comparisons, and application to Ca}

\author{Paul S. Barklem}
 \affiliation{Theoretical Astrophysics, Department of Physics and Astronomy, Uppsala University, Box 516, SE-751 20 Uppsala, Sweden}




\date{\today}

\begin{abstract}
 A theoretical method for the estimation of cross sections and rates for excitation and charge transfer processes in low-energy hydrogen atom collisions with neutral atoms, based on an asymptotic two-electron model of ionic-covalent interactions in the neutral atom-hydrogen atom system, is presented.  The calculation of potentials and non-adiabatic radial couplings using the method is demonstrated.  The potentials are used together with the multi-channel Landau-Zener model to calculate cross sections and rate coefficients.  The main feature of the method is that it employs asymptotically exact atomic wavefunctions, which can be determined from known atomic parameters.  The method is applied to Li+H, Na+H, and Mg+H collisions, and the results compare well with existing detailed full-quantum calculations.  The method is applied to the astrophysically important problem of Ca+H collisions, and rate coefficients are calculated for temperatures in the range 1000--20000~K.  
\end{abstract}

\pacs{34.50.Fa, 34.70.+e, 97.10.Ex, 97.10.Tk}
\maketitle


\section{\label{sec:intro}Introduction}

The need for data on inelastic processes due to low-energy collisions between hydrogen atoms and atoms of astrophysical interest for non-equilibrium stellar atmosphere modelling has been a long-standing problem in stellar spectroscopy \citep[e.g.][]{plaskett_interpretation_1955,steenbock_statistical_1984,Lambert1993,Barklem2011}.  
Over the last decade or so, significant progress has been made via detailed full-quantum calculations, i.e. quantum scattering calculations based on quantum chemistry calculations of the relevant molecular structure \citep[e.g.][]{Belyaev1999,Belyaev2003,Belyaev2010,Guitou2011,Belyaev2012}.  These studies have examined the inelastic processes, excitation and deexcitation
\beq
\mathrm{X}(j) + \mathrm{H} \rightleftharpoons \mathrm{X}(k) + \mathrm{H},
\label{eq:exc}
\eeq
and charge transfer (ion-pair production and mutual neutralisation)
\beq
\mathrm{X}(j) + \mathrm{H} \rightleftharpoons \mathrm{X}^+ + \mathrm{H}^- , 
\label{eq:ct}
\eeq
where X is the atom of interest, and $j$, $k$ specify different states of X.  So far such theoretical studies have covered only simpler atoms, Li, Na and Mg, and there is one experimental result for Na at intermediate energies \citep{Fleck1991}, which is well reproduced by theory \cite{Belyaev1999}.

It is well known that inelastic transitions in slow collisions may occur when potential curves approach each other and the corresponding coupling matrix element is large, and that such conditions arise in particular at avoided crossings (pseudocrossings) associated with ionic-covalent interactions \citep[e.g.][\S~3]{Bates1962}.  These crossings naturally lead to the excitation and charge transfer processes, eqns.~\ref{eq:exc} and~\ref{eq:ct}.  The detailed calculations and experimental results mentioned above, have shown this to be a key mechanism for inelastic processes in low-energy collisions with hydrogen atoms for the studied cases.  Further, charge transfer processes have been shown to generally give the largest rates, and are thus very important in astrophysical applications. In applications to stellar spectroscopy for Li and Na the charge transfer process with the largest rate has been shown to be dominant, and not very sensitive to the precision of the rate coefficient: a factor of two change leads to changes of order only one per cent in the line strengths \citep{Barklem2003b}.  In the case of Mg, charge transfer has also been shown to be very important, along with large rates for some excitation processes \cite{osorio_mg_2015}.  

As detailed calculations are extremely time consuming, a method for estimating the rates of the most important transitions with sufficient accuracy for astrophysical application is needed \cite{Barklem2011}, ideally including estimates of uncertainties.  In this work, we construct a theoretical method for estimating the relevant rates due to the mechanism described above.  The method uses an asymptotic model based on linear combinations of atomic orbitals (LCAO) for the molecular structure based on the method of Grice, Adelman and Herschbach \cite{Grice1974,Adelman1977} for treating long-range ionic-covalent interactions.  This is then coupled with a multi-channel Landau-Zener model approach to the collision dynamics.  The results are compared with the existing detailed calculations to test the approach.  Calculations are also done using semi-empirical and Landau-Herring expressions for the coupling.  These alternate approaches are found to not perform as well as the theoretical LCAO approach presented here, but provide important information that may be used to estimate the uncertainties of such model-based calculations.  We note the existence of some work using the semi-empirical and Landau-Herring approaches, namely that on alkalis \cite{Janev1978} and Al \cite{Belyaev2013,Belyaev2013a} and Si \cite{Belyaev2014}.

\section{Method}

Calculation of cross sections and rate coefficients for inelastic and charge transfer collision processes requires two main parts: a) calculation of potentials and couplings, b) solution of collision dynamics.  We describe the method for calculating potentials and couplings in \S~\ref{sect:pots}, and for performing the collision dynamics calculations to obtain the cross sections and rate coefficients in \S~\ref{sect:dynam}.   

\subsection{Potentials and couplings}
\label{sect:pots}

\subsubsection{Extended two-electron LCAO model}

The method for calculation of potentials and couplings of the diatomic system used here is an extension of the asymptotic ionic-covalent configuration mixing method proposed by Grice, Adelman and Herschbach \cite{Grice1974,Adelman1977}; hereafter GAH.  The main point of this method is that it employs asymptotically exact atomic wavefunctions, which can be determined from known atomic parameters.  This method was refined by Anstee \cite{Anstee1992}\footnote{A scanned copy of \cite{Anstee1992} may be obtained directly by contacting the author of this paper (paul.barklem@physics.uu.se).} in the context of using this theory to estimate the effect of ionic contributions to potentials in collisional broadening of the spectral lines of alkalis due to neutral hydrogen.  In particular, Anstee explicitly demonstrated that many of the integrals required in the matrix element calculations could be done fully or partially analytically.   Here, the method is refined further.  The most important improvement is that the method is extended to the case where X is a complex atom (i.e. without a spherically symmetric core and possibly with equivalent electrons), by account of coupling of angular momenta of the active electrons to those of the core electrons in the atom.   Further, correctly anti-symmetrised wavefunctions are used throughout, calculations consider all states simultaneously in solving the Schr{\"o}dinger equation, and it is demonstrated how non-adiabatic radial couplings can be calculated.   

Following GAH, we consider a diatomic system $\mathrm{X}+\mathrm{H}$ with a set of $n$ diabatic states, including a single ionic state $\Phi_1$ and a set of covalent states $\Phi_2, ..., \Phi_n$.  The corresponding adiabatic electronic wavefunctions are given by
\beq
\Psi_k = c_{1k} \Phi_1 + \sum_{j=2}^n c_{jk} \Phi_j ,
\eeq
where the coefficients $c_{ik}$ and adiabatic energies $E_k$ are found by solving the Schr{\"o}dinger equation, which is written as a generalised matrix eigenvalue equation
\beq
\mathbf{H} \mathbf{c} = E \mathbf{S} \mathbf{c}, \label{eq:genmat}
\eeq
where $\mathbf{H}$ is the electronic Hamiltonian and $\mathbf{S}$ is the overlap matrix.  

\begin{figure*}
\center
\begin{tikzpicture}[scale=5]
\path (0,0) coordinate (A);
\path (1,0) coordinate (B);
\path (0.2,0.5) coordinate (1);
\path (0.5,0.6) coordinate (2);

\path (A) ++(0,-0.01) coordinate (Alow);
\path (B) ++(0,-0.1) coordinate (Blow);

\draw [fill](A) circle (0.02);
\draw [fill](B) circle (0.03);
\draw (B) circle (0.10);
\draw [fill](1) circle (0.005);
\draw [fill](2) circle (0.005);
\draw (A) -- (B);
\draw (1) -- (2);
\draw (A) -- (2);
\draw (A) -- (1);
\draw (B) -- (2);
\draw (B) -- (1);

\node [below] at (Alow) {A};
\node [below] at (Blow) {B};
\node [above] at (1) {1};
\node [above] at (2) {2};

\node [left] at (0.1, 0.25) {$r_{1\mathrm{A}}$};
\node [left] at (0.33, 0.2) {$r_{2\mathrm{A}}$};
\node [left] at (0.7, 0.29) {$r_{1\mathrm{B}}$};
\node [left] at (0.92, 0.30) {$r_{2\mathrm{B}}$};

\node [above] at (0.35, 0.55) {$r_{12}$};
\node [above] at (0.5, 0.0) {$R$};

\draw [dashed] (0.25,0.3) circle (0.5);
\node [above] at (1.1, 0.1) {$L_c, M_{L_c}, S_c$};
\node [above] at (0.4, 0.8) {$L_{H^-}=0, M_{H^-}= 0, S_{H^-}=0$};

\end{tikzpicture}
\caption{Coordinate system for the X+H hydride system in the ionic configuration, X$^+$+H$^-$. The dashed circle encompasses the H$^-$ ion, and the full circle the bare core of atom X; in both cases the relevant quantum numbers are specified. }
\label{fig:hydrideion}
\end{figure*}
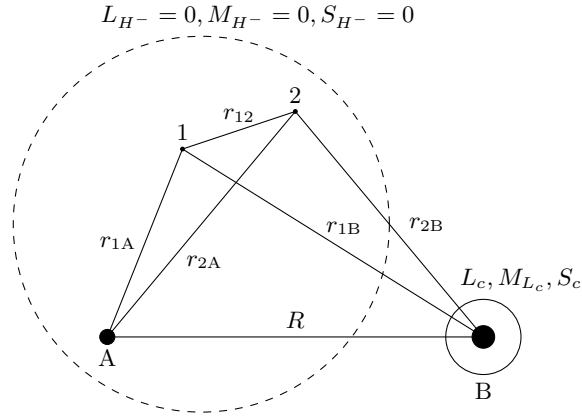

To calculate the matrices $\mathbf{H}$ and $\mathbf{S}$, we consider a model for the X+H quasimolecule with two active electrons, i.e. where only the electron on the hydrogen atom and the active electron on the atom of interest X are included explicitly.  GAH used such a model in the context of hydrogen and alkali-hydride molecules.  Here, the more general case of a possible complex atom X with a non-spherically symmetric core is considered.  The model assumes the relevant interactions at long range are dominated by a single active electron or a group of equivalent electrons.  Atom X has a charged core consisting of the nucleus and all other electrons, and the core is considered to be in some \emph{frozen} configuration and represented by some charge distribution.  The geometry for the considered system, and definition of nuclear and electronic coordinates, is shown in Figs.~\ref{fig:hydride} and~\ref{fig:hydrideion} for the covalent and ionic configurations, respectively.  The total electronic Hamiltonian is
\beq
H = - \frac{1}{2}\nabla_1^2 - \frac{1}{2}\nabla_2^2  - \frac{1}{r_{1\mathrm{A}}} - \frac{1}{r_{2\mathrm{A}}} + \frac{1}{r_{12}} + V(\vec{r}_{1\mathrm{B}}) + V(\vec{r}_{2\mathrm{B}}) + \frac{1}{R}  \label{eq:H1}
\eeq
where $V(\vec{r})$ represents a core potential, which we write  
\beq
V(\vec{r}) = - \frac{1}{r} + f(\vec{r}) ,
\eeq
where $f(\vec{r})$ is a screening function, which acts only at small $r$ and goes to zero at large $r$.  The inclusion of this screening leads to a significant increase in complexity in the calculation of matrix elements, while it does not have significant effects at large $R$.  Other approximations, such as our choice of wavefunctions, are likely to be more important, and so the screening effect is neglected ($f(\vec{r}) = 0$).

\begin{figure*}
\center
\begin{tikzpicture}[scale=5]
\path (0,0) coordinate (A);
\path (1,0) coordinate (B);
\path (0.2,0.5) coordinate (1);
\path (0.85,0.7) coordinate (2);

\path (A) ++(0,-0.01) coordinate (Alow);
\path (B) ++(0,-0.1) coordinate (Blow);

\draw [fill](A) circle (0.02);
\draw [fill](B) circle (0.03);
\draw (B) circle (0.10);
\draw [fill](1) circle (0.005);
\draw [fill](2) circle (0.005);
\draw (A) -- (B);
\draw (1) -- (2);
\draw (A) -- (2);
\draw (A) -- (1);
\draw (B) -- (2);
\draw (B) -- (1);

\node [below] at (Alow) {A};
\node [below] at (Blow) {B};
\node [above] at (1) {1};
\node [above] at (2) {2};

\node [left] at (0.1, 0.25) {$r_{1\mathrm{A}}$};
\node [left] at (0.39, 0.2) {$r_{2\mathrm{A}}$};
\node [left] at (0.7, 0.29) {$r_{1\mathrm{B}}$};
\node [left] at (0.92, 0.35) {$r_{2\mathrm{B}}$};

\node [above] at (0.5, 0.6) {$r_{12}$};
\node [above] at (0.5, 0.0) {$R$};

\draw [dashed] (1.0,0.32) circle (0.5);
\node [above] at (1.1, 0.1) {$L_c, M_{L_c}, S_c$};
\node [above] at (1.5, 0.7) {$L_A, M_{L_A}, S_A$};

\draw [dashed] (0.0,0.32) circle (0.5);
\node [above] at (-0.3, 0.7) {$L_H=0, M_H=0, S_H=1/2$};

\end{tikzpicture}
\caption{Coordinate system for the X+H hydride system in the covalent configuration.  The hydrogen atom is centred on A and the target atom X centred on B.  The atom X is considered as a charged core with a single active electron.  The two electrons considered explicitly are labelled 1 and 2.  The dashed circles delineate the two atoms, and their relevant quantum numbers.}
\label{fig:hydride}
\end{figure*}
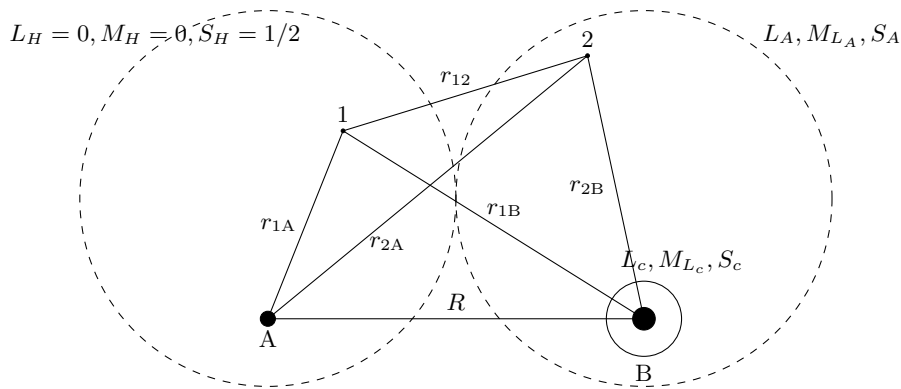

For the ionic-covalent configuration mixing problem it is convenient to partition the Hamiltonian in two ways.  The covalent partitioning corresponding to Fig.~\ref{fig:hydride} is
\beq
H = H_\mathrm{cov} - \frac{1}{r_{1\mathrm{B}}} - \frac{1}{r_{2\mathrm{A}}} + \frac{1}{R} + \frac{1}{r_{12}},
\eeq
where
\beq
H_\mathrm{cov} = - \frac{1}{2}\nabla_1^2 - \frac{1}{2}\nabla_2^2  - \frac{1}{r_{1\mathrm{A}}} - \frac{1}{r_{2\mathrm{B}}}  , \label{eq:Hlast}
\eeq
corresponding to the electronic Hamiltonian of the two atoms at infinite separation.  The ionic partitioning corresponding to Fig.~\ref{fig:hydrideion} is:
\beq
H = H_\mathrm{ion} - \frac{1}{r_{1\mathrm{B}}} - \frac{1}{r_{2\mathrm{B}}} + \frac{1}{R} ,
\eeq
where
\beq
H_\mathrm{ion} = - \frac{1}{2}\nabla_1^2 - \frac{1}{2}\nabla_2^2  - \frac{1}{r_{1\mathrm{A}}} - \frac{1}{r_{2\mathrm{A}}} + \frac{1}{r_{12}},
\eeq
and is the electronic Hamiltonian of the H$^-$ ion, i.e. when both electrons are located on A.  This partitioning allows matrix elements of $H_\mathrm{cov}$ and $H_\mathrm{ion}$ to be related to atomic energies, such that only the remaining interaction terms need to be calculated.

\begin{widetext}
The anti-symmetrised wavefunction for the diatom at large internuclear distance is written in terms of atomic one-electron wavefunctions for the active electron on B and the electron on the ground state neutral hydrogen atom, which in $LS$ coupling and neglecting spin-orbit coupling can be written (see \cite[ch. 5,][]{sobelman_atomic_1979} and \cite[ch. 3,][]{Nikitin1984})
\bea
\label{eqn:cov}
\Phi_j (L, \tilde{\Lambda}, S, M_S) & = & \hat{A} \sum_{S_c L_c} G_{S_c L_c}^{S_A L_A} \sum_{M_{S_A} \sigma M_{L_A}} 
\left[ \begin{array}{ccc}
S_A & 1/2 & S \\
M_{S_A} & \sigma & M_{S} \\
\end{array} \right] 
\left[ \begin{array}{ccc}
L_A & 0 & L \\
M_{L_A} & 0 & \tilde{\Lambda} \\
\end{array} \right] \n &&
\sum_{M_{S_c} \mu M_{L_c} m}
\left[ \begin{array}{ccc}
S_c & 1/2 & S_A \\
M_{S_c} & \mu & M_{S_A} \\
\end{array} \right] 
\left[ \begin{array}{ccc}
L_c & l & L_A \\
M_{L_c} & m & M_{L_A} \\
\end{array} \right]
\psi(L_c,M_{L_c},S_c,M_{S_c}) \psi^\mathrm{A}_{1s\sigma} \psi^\mathrm{B}_{j\mu} .
\eea 
Note, the symbol $\tilde{\Lambda}$ is the projection of the orbital angular momentum $L$ along the internuclear axis, and the tilde distinguishes it from the absolute value $\Lambda = |\tilde{\Lambda}|$ usually used in denoting the molecular term.  The function $\psi^\mathrm{A}_{1s\sigma}$ is the hydrogen $1s$ wavefunction with spin projection quantum number $\sigma$ located on A, and $\psi^\mathrm{B}_{j\mu}$ represents the wavefunction of the valence electron located on B, with $j$ being an index corresponding to the orbital of interest and thus to the quantum numbers $nlm$, and $\mu$ the spin projection quantum number.  The function $\psi(L_c,M_{L_c},S_c,M_{S_c})$ is the wavefunction of the core electrons on B.  $\hat{A}$ is the anti-symmetrisation operator (containing also normalisation factors), $G$ is the coefficient of fractional parentage, and the bracketed symbols are the Clebsch-Gordon coefficients, such that $\left[\begin{array}{ccc} j_1 & j_2 & J \\ m_1 & m_2 & M \\ \end{array} \right] = \la (j_1 j_2)m_1 m_2|JM\ra$. In the ionic state, the two electrons on A, the H$^-$ ion, must form a singlet state, and the wavefunction is written
\bea
\Phi_{1} (L, \tilde{\Lambda}, S, M_S) & = & \hat{A}
\sum_{\mu}
\left[ \begin{array}{ccc}
1/2 & 1/2 & 0 \\
\mu & -\mu & 0  \\
\end{array} \right] 
 \delta_{\tilde{\Lambda} M_{L_c}} \delta_{S S_c}\delta_{M_S M_{S_c}}  \psi(L_c,M_{L_c},S_c,M_{S_c}) \psi^\mathrm{A}_{1s\sigma} \psi^\mathrm{A}_{LR\mu} .
\eea
\end{widetext}
The function $\psi^\mathrm{A}_{LR\mu}$ is a long-range approximation to the H$^-$ function with spin projection quantum number $\mu$ ($=-\sigma$), and the spatial part is specified below.  Note, the quantum number $L$ is only good asymptotically.  In any case, since both the ground state hydrogen atom and the H$^-$ ion have zero orbital angular momentum, we have $L=L_c$ for the ionic state $\Phi_1$, and $L=L_A$ for the covalent state $\Phi_j$, and thus it does not enter the calculations, and is hereafter omitted.  

The one-electron functions are written in terms of spatial and spin functions such that $\psi_{nlm\sigma}= \varphi_{nlm}(\vec{r}) \chi_{\frac{1}{2},\sigma}$, where $\chi_{\frac{1}{2},\sigma}$ is the spin function with projection quantum number $\sigma$.  We define the hydrogen $1s$ state spatial function
$
\varphi^\mathrm{A}_{1s}(r) = \varphi^\mathrm{A}_{0}(r) = e^{-r}/\sqrt{\pi},
$
noting that the subscript 0 will be sometimes used as shorthand for 1s. The function $\varphi^\mathrm{A}_{LR}$ is the approximate long-range H$^-$ spatial function, here following GAH
\beq
\varphi^\mathrm{A}_{LR}= \left\{ \begin{array}{cl} 
        Ne^{-\gamma r}/r & r \ge r_0 \\
		0                        & r < r_0
		\end{array} \right. ,
\eeq
where $N$ = .223106, $\gamma $ = .2355885, for $r$ in atomic units, which were adjusted to match the variational wavefunction of \cite{pekeris_ground_1958}.  Following \cite{Anstee1992}, we choose $r_0$ = .601324404~a.u., which gives a correctly normalised $\varphi^\mathrm{A}_{LR}$.  Note, the subscript $L$ will be used as shorthand for $LR$ as required.  The function $\varphi^\mathrm{B}_j$ is the spatial wavefunction of the active electron on B 
\beq
\varphi^\mathrm{B}_{j}(\vec{r}) = \varphi^\mathrm{B}_{nlm}(\vec{r}) =  \frac{P_{nl}(r)}{r} Y_l^m(\Omega_r). 
\eeq
The radial wavefunctions $P_{nl}(r)$ are calculated numerically, and this will be described further below.

Since the ionic function $\Phi_1$ forms a singlet state for the two electrons forming the H$^-$ ion, and the considered Hamiltonian is purely electrostatic, the Hamiltonian and overlap matrix elements will only be non-zero between the ionic state and covalent states of the same symmetry i.e. with the same quantum numbers $\tilde{\Lambda}, S, M_S$.  As the Hamiltonian does not involve the core electrons, they are purely spectators, the Hamiltonian and overlap matrix elements will only be non-zero between the ionic state $\Phi_1$, and covalent states $\Phi_j$ that a) have the same core, b) also form a singlet state from the two active electrons, and c) have the valence electron on B with projection of orbital angular momentum quantum number $m=0$.  The function $\Phi_j$ in eqn~\ref{eqn:cov} is written with the coupling order $(\vec{J_c} + \vec{J_v}) + \vec{J_0}$, where subscript $v$ refers to the valence electron, which were above labelled $j$ when on B and with LR when on A, and recalling that the label $0$ is used for the hydrogen $1s$ electron on A.  This is necessary, since the angular momentum of the atom X arises from $\vec{J_A} = \vec{J_c} + \vec{J_v}$.  However, for the ionic state, the coupling order $\vec{J_c} + (\vec{J_v} + \vec{J_0})$ is required as the valence and hydrogen $1s$ electrons electrons couple to give the H$^-$ ion.  These electrons form a singlet state, and only components of the covalent state where $\vec{S_j} + \vec{S_0}$ also form a singlet state lead to non-zero matrix elements.  Thus, we rewrite the (terms with appropriate symmetry of the) covalent state with this coupling order via re-coupling, and requiring that $\vec{S'} = \vec{S_v} + \vec{S_0}$ forms a singlet state $S'=0$, and that $m=0$, one obtains: 
\begin{widetext}
\bea
\Phi_j (\tilde{\Lambda}, S, M_S) & = & \hat{A} \sum_{S_c L_c} G_{S_c L_c}^{S_A L_A} \sum_{M_{S_A} \sigma M_{L_A}} 
\delta_{\tilde{\Lambda}M_{L_c}} \delta_{S S_c} \delta_{M_{S}M_{S_c}}  \n &&
(-1)^{3(Sc+\frac{1}{2}+S_A)} \sqrt{\frac{2S_A+1}{2(2S_c+1)}}
\left[ \begin{array}{ccc}
L_c & l & L_A \\
\tilde{\Lambda} & 0 & \tilde{\Lambda}  \\
\end{array} \right] 
\psi(L_c,M_{L_c},S_c,M_{S_c}) \psi^\mathrm{A}_{1s\sigma} \psi^\mathrm{B}_{j\mu} .
\eea
\end{widetext}

Since the core is a spectator only, this means that the required Hamiltonian and overlap matrix elements can be written in terms of two-electron matrix elements.  It is convenient to separate out effects of the coupling between the core and the active electrons.  If the core is neglected, using NC (``neglecting core'') to denote such functions and matrix elements, one obtains for the ionic state:
\beq
\Phi_1^\mathrm{NC} = \frac{1}{2\sqrt{(1+S_{0L}^2)}} ( \varphi^\mathrm{A}_{1s} \varphi^\mathrm{A}_{LR} + \varphi^\mathrm{A}_{LR} \varphi^\mathrm{A}_{1s}) \chi_{00}, \label{eq:ionicwf}
\eeq
where $\chi_{00}$ is the two-electron spin function for $S'=0$, $M_S'=0$.  For the covalent function, the singlet case is:
\beq
\Phi_j^\mathrm{NC} = \frac{1}{2\sqrt{(1+S_{0j}^2)}} ( \varphi^\mathrm{A}_{1s} \varphi^\mathrm{B}_j + \varphi^\mathrm{B}_j \varphi^\mathrm{A}_{1s})  \chi_{00}, \label{eq:covwf}
\eeq
which is the Heitler-London function, and roughly equivalent to the functions of GAH and Anstee, except that our ionic function is correctly anti-symmetrised.  In writing these functions, we have used the convention than for any pair of orbitals the first orbital refers to the electron with label 1, and the second orbital refers to the electron with label 2, e.g. $\varphi^\mathrm{A}_{1s} \varphi^\mathrm{B}_j = \varphi^\mathrm{A}_{1s}(r_{1\mathrm{A}}) \varphi^\mathrm{B}_j(r_{2\mathrm{B}})$.  $S_{0j} = \la \varphi^\mathrm{A}_{1s} | \varphi^\mathrm{B}_j \ra$, $S_{0L} = \la \varphi^\mathrm{A}_{1s} | \varphi^\mathrm{A}_{LR} \ra$ and $S_{jL} = \la \varphi^\mathrm{B}_j | \varphi^\mathrm{A}_{LR} \ra$ are the one-electron overlaps between the atomic wavefunctions.  Note, the one-electron and multi-electron overlaps are both denoted by $S$, but can be distinguished by the presence of subscripts corresponding to one-electron functions on A, i.e. 0 and $L$.  The cases of one-electron overlaps on B, $S_{jk}= \la \varphi^\mathrm{B}_j | \varphi^\mathrm{B}_k \ra$, and multi-electron overlaps $S_{jk} =  \la \Phi_j | \Phi_k \ra$ need not be distinguished, since these functions are always assumed orthonormal, and thus have the same values. 

Diagonal matrix elements are unaffected by the other electrons and therefore $H_{11} = H_{11}^\mathrm{NC}, H_{jj} = H_{jj}^\mathrm{NC}$, and overlaps are unity in both cases, $S_{11}=S_{jj}=1$.  The off-diagonal elements between the ionic and other covalent states provide the mechanism for coupling, and for a chosen core we obtain:
\begin{widetext}
\beq
H_{1j}(\tilde{\Lambda}, S) =  \delta_{\tilde{\Lambda} M_{L_c}} \delta_{S S_c} \sqrt{N_\mathrm{eq}} \, G^{S_A L_A}_{S_c L_c} (-1)^{3(Sc+\frac{1}{2}+S_A)} \sqrt{\frac{2S_A+1}{2(2S_c+1)}} \left[ \begin{array}{ccc}
L_c & l & L_A \\
\tilde{\Lambda} & 0 & \tilde{\Lambda}  \\
\end{array} \right] 
 \times H_{1j}^\mathrm{NC}
\eeq
\end{widetext}
where $N_\mathrm{eq}$ denotes the number of equivalent valence electrons on B, and $H_{1j}^\mathrm{NC} = \la \Phi_1^\mathrm{NC} | H | \Phi_j^\mathrm{NC} \ra$.  The label $M_S$ has been dropped, since as is well known it does not affect the molecular energy, and leads only to $2S+1$ degenerate states.  For brevity we write this as
\beq
H_{1j}(\tilde{\Lambda}, S) =  C \times H_{1j}^\mathrm{NC}
\eeq
and similarly
\beq
S_{1j}(\tilde{\Lambda}, S) =  C \times S_{1j}^\mathrm{NC}
\eeq
where $S^\mathrm{NC}_{1j} = \la \Phi_1^\mathrm{NC} | \Phi_j^\mathrm{NC} \ra$.  For the case of a spherical core one finds $C=1$, and thus attains the case closest to the work of GAH and Anstee on alkali-hydride and hydrogen molecules.  For a filled valence shell, one obtains $C = 1/\sqrt{2}$.  After deriving this expression, we noted that a more general expression had been derived by Smirnov \cite{smirnov_asymptotic_1973}, available only in Russian, but which is reproduced as equations 3.26-3.27 of \cite{Chibisov1988}.  Our expression is the specific case of one of the atoms being a ground state hydrogen atom.   Note, in some cases it may be necessary to deal with multiple open shells, and thus calculate the coefficient of fractional parentage for mixed configurations.  This can be achieved with the formulae given by \cite{Kelly1959}.

\subsubsection{Practical implementation of the model}

The task is now to calculate the matrix elements $H_{ij}^\mathrm{NC}$ and $S_{ij}^\mathrm{NC}$ using the two-electron wavefunctions in terms of one-electron orbitals defined above.  The off-diagonal matrix elements involving only covalent states are always assumed asymptotically small, and thus are set to zero, i.e. $H_{jk}=0, S_{jk}=0$.  Thus, the matrices are ``arrowhead'' in form.  The partitioning of the Hamiltonian allows the remaining matrix elements to be written in terms of atomic energies and one-electron matrix elements of the interaction terms, and the results are given in Appendix~\ref{app:twomat}.  The results for the overlaps are also given there.  Expressions for the required one-electron matrix elements were derived analytically as far as possible with the aid of Mathematica by Anstee, and can be found in \cite{Anstee1992}; see Appendix~\ref{app:onemat} for an overview of the required matrix elements and brief discussion.  For matrix elements not involving $\varphi^\mathrm{B}_j$, the matrix elements can be derived fully analytically.  Those involving $\varphi^\mathrm{B}_j$ require a single numerical integral over the radial part of the wavefunction on B, $P_{nl}(r)$.  The radial wavefunctions $P_{nl}(r)$ are calculated numerically using the scaled Thomas-Fermi-Dirac method \citep{Warner1968c}.   This method employs a radially symmetric core potential $V(\vec{r}) = V_{TFD}(r)$ in the atomic Hamiltonian, semi-empirically scaled to produce the correct energy eigenvalue, and though the core is neglected in the total electronic Hamiltonian for the quasi-molecule, this simple model of the core is included here as it leads to no additional effort compared to Coulomb wavefunctions as used by GAH.  Note, in performing this calculation the correct energy eigenvalue is the binding energy of the considered active electron.  This is the difference between state energy $E$ and the series limit $E_{lim}$ corresponding to the core configuration (i.e. the appropriate state of X$^+$ corresponding to the core).   
 
 Given the matrices, $\mathbf{H}$ and $\mathbf{S}$, corresponding to matrix elements $H_{ij}$ and $S_{ij}$, eqn~\ref{eq:genmat} can then be solved and adiabatic wavefunctions and potentials obtained; i.e. $| \Psi_j \ra$ via the matrix $\mathbf{c}$, and potentials $E_j(R)$ solved for various internuclear distances $R$.  Full-quantum dynamical calculations require not only potential energies, but also non-adiabatic radial couplings $\la \Psi_i | \frac{\partial}{ \partial R} | \Psi_j \ra$.  A method for calculating the radial couplings from this model is described in Appendix~\ref{app:radcoup}.  Note, radial couplings are not employed in the calculations done here, but the method is presented as it could be useful for combining the asymptotic model presented here with quantum chemistry results in quantum scattering calculations.   Computer codes to generate the wavefunctions $P_{nl}(r)$, compute the matrices, solve eqn~\ref{eq:genmat} and calculate the radial couplings have been written in Fortran 95, using codes written by \cite{Anstee1992} in Fortran 77 for the hydrogen and alkali-hydride cases as a starting point.   Given the matrices $\mathbf{H}$ and $\mathbf{S}$ in the diabatic basis, equation~\ref{eq:genmat} can be solved with standard numerical methods; we use \texttt{DSYGV} from \texttt{LAPACK}.  The problem is solved on an adaptive grid of $R$ in order to sufficiently resolve narrow crossing regions.  It was found in \cite{Adelman1977} that the matrices could be truncated to omit weakly coupled states with only small effects.  Thus, GAH and Anstee considered only two- and three-state treatments, which can be easily solved analytically via the usual secular determinantal equation.  However, with modern computers and widely available numerical libraries such as that mentioned above, we can easily include all states together in a single calculation.

\subsubsection{Alternate models: semi-empirical and Landau-Herring estimates}

We also consider alternate methods for estimating the coupling $H_{1j}^\mathrm{NC}$ at avoided crossings, in addition to the theoretical LCAO model presented here.  In particular we calculate for three models considered by \cite{MiyanoHedberg2014}, namely a semi-empirical method, and two models based on the Landau-Herring method \cite{Herring1962,Herring1964,landau_quantum_1965}.  Our implementation of these models follows \cite{MiyanoHedberg2014}, and that paper can be consulted for a more thorough discussion of the physics.  The semi-empirical expression for the coupling is given by Olson \cite{olson_absorbingsphere_1972}, building on earlier work.  The two expressions for the coupling from the Landau-Herring method, are those of Smirnov \cite{smirnov_formation_1965,smirnov_negativeion_1967} and Janev \cite{Janev1976}.  The expressions differ due to choice of the hyper-surface demarcating the regions of influence of the two atoms.  We used these expressions directly to calculate the $H_{1j}^\mathrm{NC}$ parameter at the avoided crossing position given from the asymptotic LCAO model (see below).  For ease of discussion, we will label the four models considered here as LCAO, Semi-emp, LH-S, and LH-J.
\subsection{Collision dynamics}
\label{sect:dynam}

The collision dynamics are treated by the multi-channel Landau-Zener (LZ) model as presented in \cite{Belyaev1993,Belyaev2003}, which considers a single ionic state, with label 1, crossing a series of covalent states, labels $2,...,n$.  The LZ model provides a relatively simple way to estimate the transition probability at a given avoided crossing, via a linear two-state description.  The model is described in detail in many places; our discussion here is based primarily on the formulation of \cite{Nikitin1984}.

The Hamiltonian in an orthonormal diabatic representation such that
\beq
\mathbf{H}^\mathrm{d} \mathbf{c} = E \mathbf{c}  \label{eq:genmat_ortho}
\eeq
is formulated via a linear two-state model
\beq
\mathbf{H}^\mathrm{d} = \left(
\begin{array}{cc}
U_c - F_1(R-R_c) & a \\
a & U_c - F_j(R-R_c) \\
\end{array}
\right),
\eeq
where $R_c$ is the crossing point, $U_c$ is the potential energy at the crossing point, and $a$ the coupling, assumed constant.  For the case where the crossing is far from the turning point, and assuming uniform nuclear motion such that $X(t) = R-R_c = v_c t$ where $v_c$ is the radial velocity at the crossing point, the single passage transition probability is
\beq
p_{1j} = \exp \left( \frac{-2\pi a^2}{ |\Delta F| v_c} \right),
\eeq
where $\Delta F = F_1 - F_j$.   Thus, the paramters $R_c$, $a$, $\Delta F$ and $U_c$ need to be derived from the molecular structure calculations.  In our case, this can be done in two ways:  directly from the Hamiltonian and overlap matrices in the non-orthogonal diabatic representation, or from the potentials in the adiabatic representation.  These two methods are described below, followed by a discussion of the choice of a final set of parameters.

\subsubsection{Diabatic representation}

The diabatic representation used in this work is non-orthogonal and thus the above cannot be applied directly.  However, the relationship between the two representations is relatively simple to derive.  The Schr{\"o}dinger equation for the non-orthogonal case, eqn~\ref{eq:genmat}, can be rewritten as
\beq
\mathbf{S}^{-1} \mathbf{H} \mathbf{c} = E  \mathbf{c} \label{eq:genmat2}
\eeq
and so 
\beq
\mathbf{H} = \mathbf{S} \mathbf{H}^\mathrm{d} .
\eeq
We consider the two-state case, where the matrix elements pertaining to the ionic state $\Phi_1$ and a given covalent state $\Phi_j$ are extracted from the total matrices.  We write the two-state overlap matrix as 
\beq
\mathbf{S} = \left(
\begin{array}{cc}
1 & {S}_{1j}  \\
{S}_{1j} & 1 \\
\end{array}
\right).
\eeq
We then obtain
\beq
\mathbf{H} = \left(
\begin{array}{cc}
a {S}_{1j} + U_c - F_1 X & a + {S}_{1j}(U_c - F_j X) \\
a + {S}_{1j}(U_c - F_1 X) & a {S}_{1j} + U_c - F_j X \\
\end{array}
\right) .
\eeq
By equating this to the matrix elements of $\mathbf{H}$ and solving the equations we can find expressions for the required LZ model parameters in terms of these matrix elements obtained in the non-orthogonal basis.  We find
\beq
H_{11} - H_{jj} = - \Delta F \, X = -(F_1-F_j) (R-R_c).
\eeq
Thus, $R_c$ and $\Delta F$ are the same in both representations.  The coupling in the orthonormal diabatic representation is given in terms of the matrix elements in the non-orthogonal representation by
\beq
a = (H_{1j} - H_{11} S_{1j}) / (1-S_{1j}^2).
\eeq
and the equivalent potential energy at the crossing by
\beq
U_c = (H_{11} - H_{1j} S_{1j}) / (1-S_{1j}^2),
\eeq
which will be required below to calculate the velocity.  Noting that $a = \Delta E /2$ , where $\Delta E$ is the splitting between adiabatic curves at the avoided crossing (see below), the expression for the coupling agrees with eqn.~1 of Adelman and Herschbach \cite{Grice1974}.

\subsubsection{Adiabatic representation}

The LZ model parameters can also be related to the adiabatic potentials $E$ via the solution of eqn.~\ref{eq:genmat_ortho}. The splitting between the adiabatic potentials is given by the well-known result \cite[e.g.][Chap. 8]{Nikitin1984}
\beq
\Delta E (X) = ( \Delta F^2 X^2 + 4 a^2) ^{1/2}.
\eeq
The crossing distance $R_c$, where $X=0$, is the distance with the minimum splitting $\Delta E$.  The coupling $a$ can be obtained from the splitting at this point simply via $a = \Delta E (R_c) /2$.  The slope difference $\Delta F$ can be obtained from fitting to $\Delta E$ for small $X$.  However, \cite{belyaev_nonadiabatic_2011} have shown that $\Delta F$ can be more elegantly written in terms of the splitting $\Delta E$ and its second derivative $\Delta E^{''}$, via
\beq
\Delta F = \sqrt{\Delta E \Delta E^{''} } .
\eeq

\subsubsection{Calculation of LZ parameters}

It has been shown by \cite{Belyaev2010} that the off-diagonal coupling element in a $n$-state diabatic basis can differ markedly from that in the two-state basis, particularly for crossings at short internuclear distances where many diabatic states may interact. To most correctly capture the dynamics, the LZ parameters should be as close as possible to the two-state representation, which can be achieved by deriving the LZ parameters from the adiabatic potentials.  Thus, in the cases of crossings at relatively short internuclear distances the LZ parameters are best estimated from the adiabatic potentials.  On the other hand, calculation of the LZ parameters from the adiabatic potentials has drawbacks for rather narrow crossing regions at large internuclear distance.  The calculation of splittings and the second derivative may be susceptible to numerical errors due to limitations of the grids of calculated $R$.  For well localized crossings at large distance, differences between the two-state and $n$-state representations become small, allowing the LZ parameters to be safely extracted directly from the diabatic calculations.  This suggests a hybrid approach where LZ parameters from the adiabatic potentials are used for crossings at small $R$ and parameters from the diabatic representation at large $R$.

Thus, to derive the LZ parameters, they are first calculated from the diabatic representation.  This allows us to use these parameters for long-range crossings, as well as providing a first estimate of the ionic crossing location $R_c$, which is very easily and uniquely found in the diabatic representation.  A second set of parameters is then derived from the adiabatic potentials, where the diabatic value of the crossing distance $R_c$ is used as an initial guess.  As adiabatic potentials may show more than one minimum in the splitting, this helps to ensure the automated algorithms find the correct avoided crossing.

We finally adopt the adiabatic parameters for crossings at $R < 50$ a.u., while we adopt the diabatic parameters at larger distances.   This switch-over point was determined empirically from examining differences between the two sets of LZ parameters in various test calculations, all of which show the expected differences at small and large $R$.  However, in the intermediate region, roughly $ 20 < R < 70$~a.u., the two methods are typically in good agreement.

\subsubsection{Calculation of cross sections and rate coefficients}

The cross section, for collision energy $E$, is computed as a sum over partial waves
\beq
\sigma_{if}(E) = \frac{\pi \hbar^2 p_i^\mathrm{stat} }{2 \mu E} \sum_{J=0}^\infty P_{if}(J,E) \times (2J+1),
\eeq   
where $p_i^\mathrm{stat}= 1/g_i$ is the statistical probability for population of the initial channel $i$; $g_i$ is the statistical weight of the channel.  $P_{if}(J,E)$ is the multi-channel transition probability, calculated according to expressions given in \cite{Belyaev1993,Belyaev2003} employing the individual LZ crossing probabilities $p_{1j}$ calculated as detailed above, where the velocity is given by
\beq
v_c = \sqrt{\frac{2}{\mu} \left( E + U_i(R=\infty) - U_c - \frac{J(J+1)\hbar^2}{2\mu R_c^2} \right) },
\eeq
where $U_i(R=\infty)$ is the potential energy of the initial channel at infinite separation.

To calculate the \emph{complete} cross section and rate coefficients for processes of a given atom X, the above method is applied for all relevant cores and all possible symmetries defined by quantum numbers $\Lambda, S$.  In the case of $\Sigma$ states, $\Lambda=0$, the reflection symmetry ($+$/$-$) must also be considered.  As a starting point for any calculation, a list of states of atom X to be considered is compiled, along with a list of states of ion X$^+$, which defines the possible core states to be considered.  For all these states, the required quantum numbers are compiled, along with the energy of the state $E_j$, $N_{eq}$, and which core corresponds to the given state as well as the series limit energy corresponding to this core $E_{lim}$.  Note, a given state $^{2S+1}L$ may have more than one possible core in the case that there is more than one possible choice of valence electron, or in the case of equivalent electrons with fractional parents.  All possible molecular terms ($^{2S+1}\Lambda^{+/-}$) are then derived from the asymptotic atomic states (e.g. \cite{Herzberg1950, landau_quantum_1965}).  One then calculates for each possible symmetry of the considered cores, noting that only these symmetries lead to ionic-covalent coupling.  The symmetries derived are used to calculate the appropriate statistical probabilities for the initial channel $p_i^\mathrm{stat}$.  Finally, results are summed over all possible cores and symmetries to give the final cross sections, which are then integrated over Maxwellian velocity distributions to obtain the rate coefficients.  This will be illustrated in the next section by application to Ca+H collisions.

The handling of data, including the calculation of relevant symmetries, and passing to relevant Fortran codes for calculation of the potentials and dynamics is controlled by a code written in Interactive Data Language (IDL).

\section{Results}

In this section, the results of our calculations with the new LCAO model are presented, as well as those with the alternate models.  First, in \S~\ref{sect:compres} the results of the model approaches are compared with detailed full-quantum calculation results.  This allows us to evaluate the success of the model approaches.  Second, in \S~\ref{sect:Ca_res} the results for Ca+H are presented, including detailed description of the input data, which demonstrates how the method is used in practice.

\subsection{Comparison with full-quantum calculations involving simple atoms: Li, Na, Mg}
\label{sect:compres}

As mentioned in the introduction, full-quantum calculations, i.e. quantum scattering calculations based on quantum chemistry calculations of the relevant molecular structure, have been performed for Li+H \citep{Belyaev2003,Barklem2003b}, Na+H \citep{Belyaev2010,Barklem2010}, and Mg+H \citep{Guitou2011,Belyaev2012,Barklem2012}.  These calculations are based on data for the Li+H \citep{Croft1999a,Gadea1993,boutalib_abinitio_1992}, Na+H \citep{dickinson_initio_1999}, and Mg+H \citep{Guitou2010} molecules from quantum chemistry type calculations, including potentials and couplings.  The quantum scattering calculations, at least for the low-lying states, are done by the $t$-matrix reprojection method \citep{belyaev_dependence_2002,belyaev_revised_2010,grosser_approach_1999,belyaev_electron_2001,Belyaev2009}, which solves the so-called electron translation problem \citep{bates_electron_1958}.  For more details of the calculations the relevant papers should be consulted.  

\begin{figure*}
\centering
\begin{overpic}[width=0.32\textwidth]{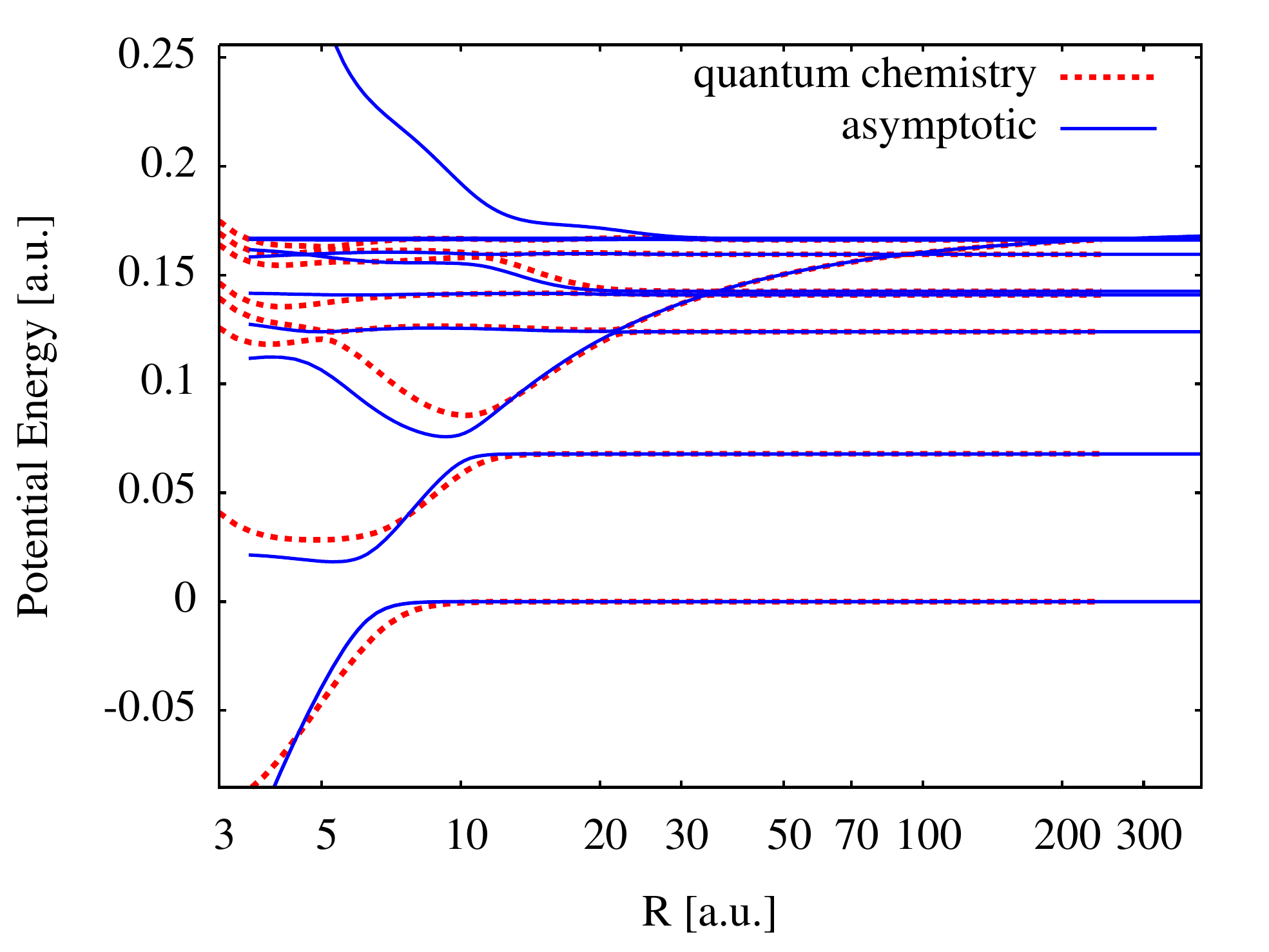}\put(70,15){Li+H}\end{overpic}
\begin{overpic}[width=0.32\textwidth]{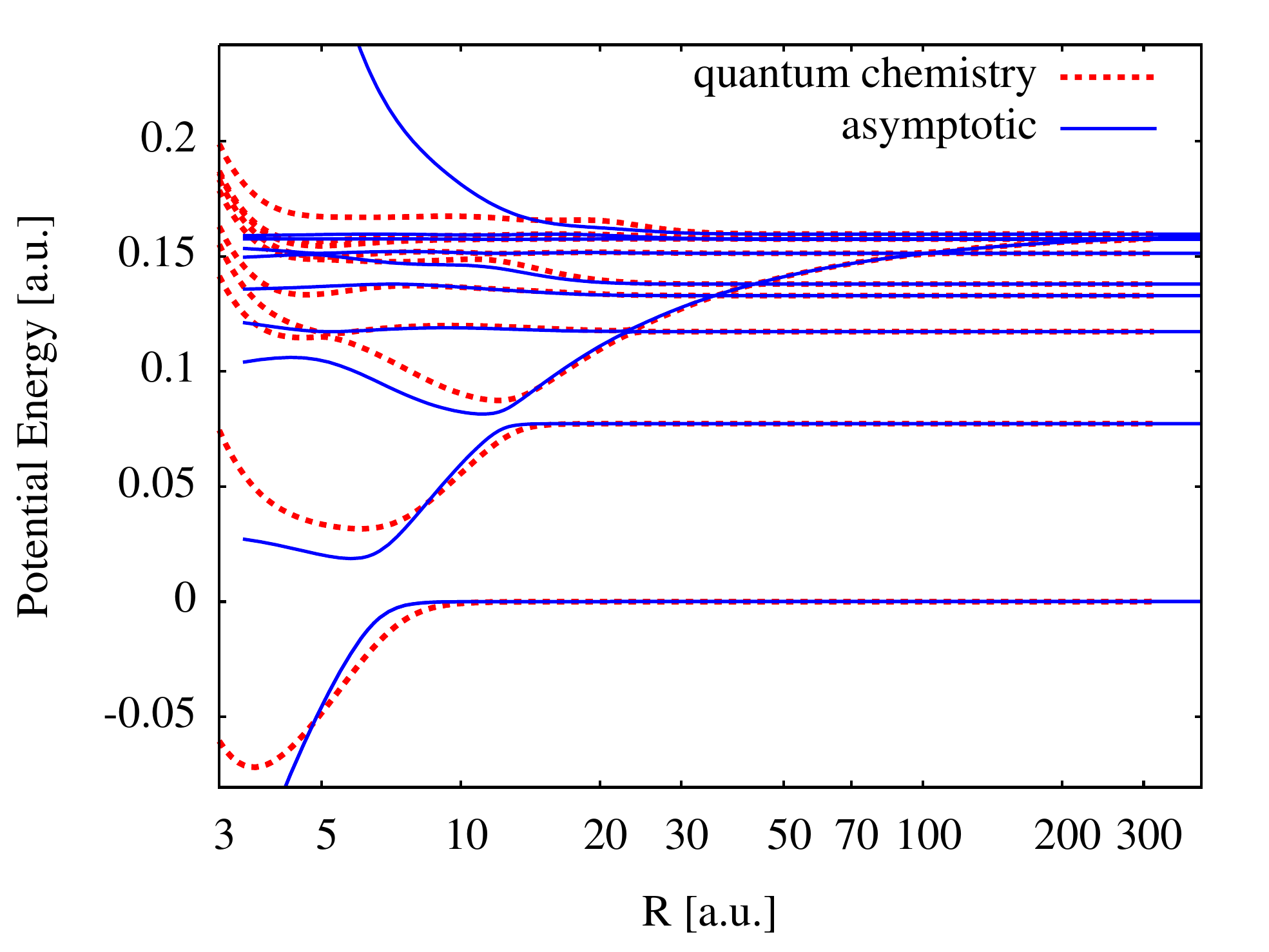}\put(70,15){Na+H}\end{overpic}
\begin{overpic}[width=0.32\textwidth]{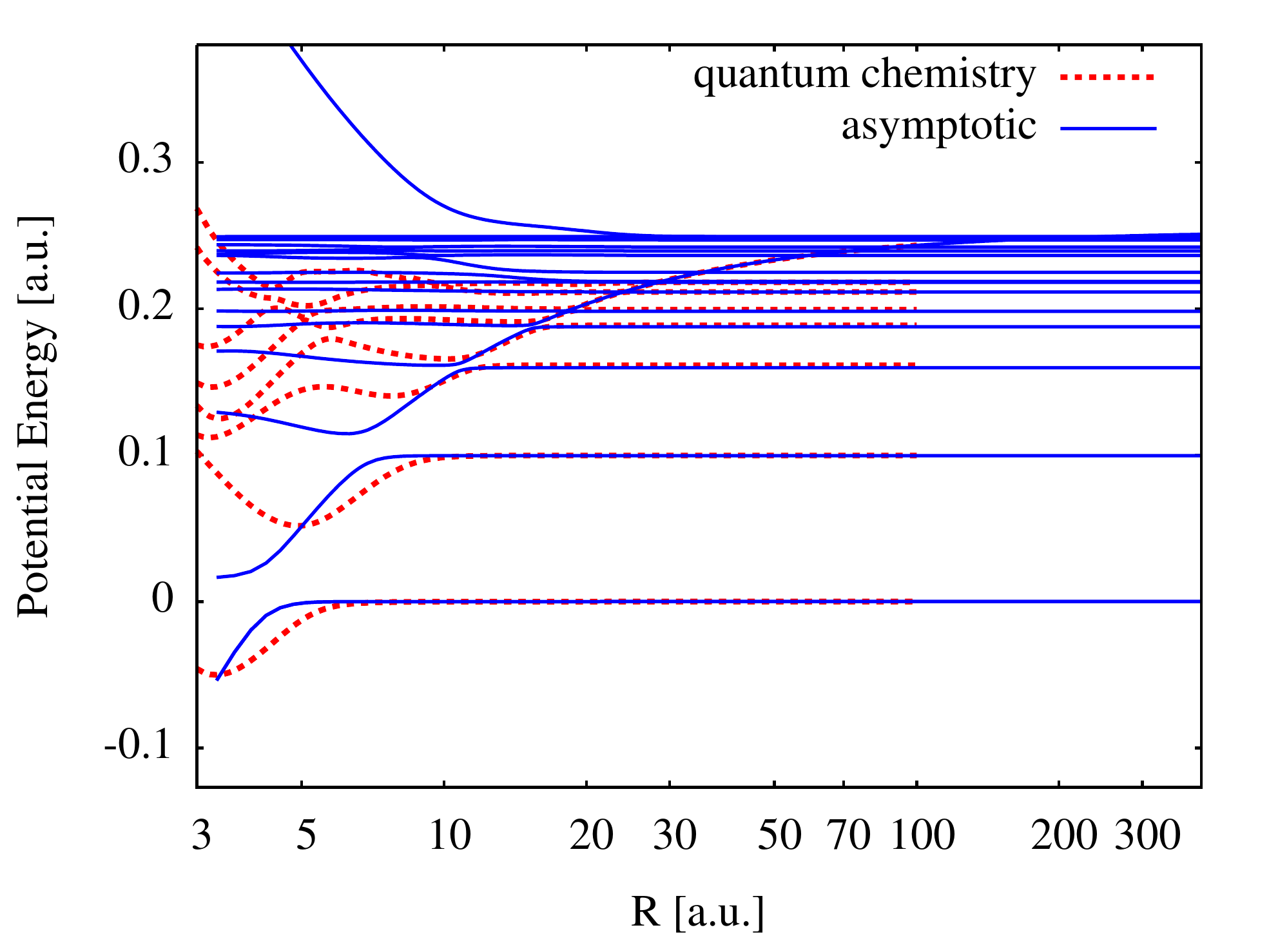}\put(70,15){Mg+H}\end{overpic}
\includegraphics[width=0.32\textwidth]{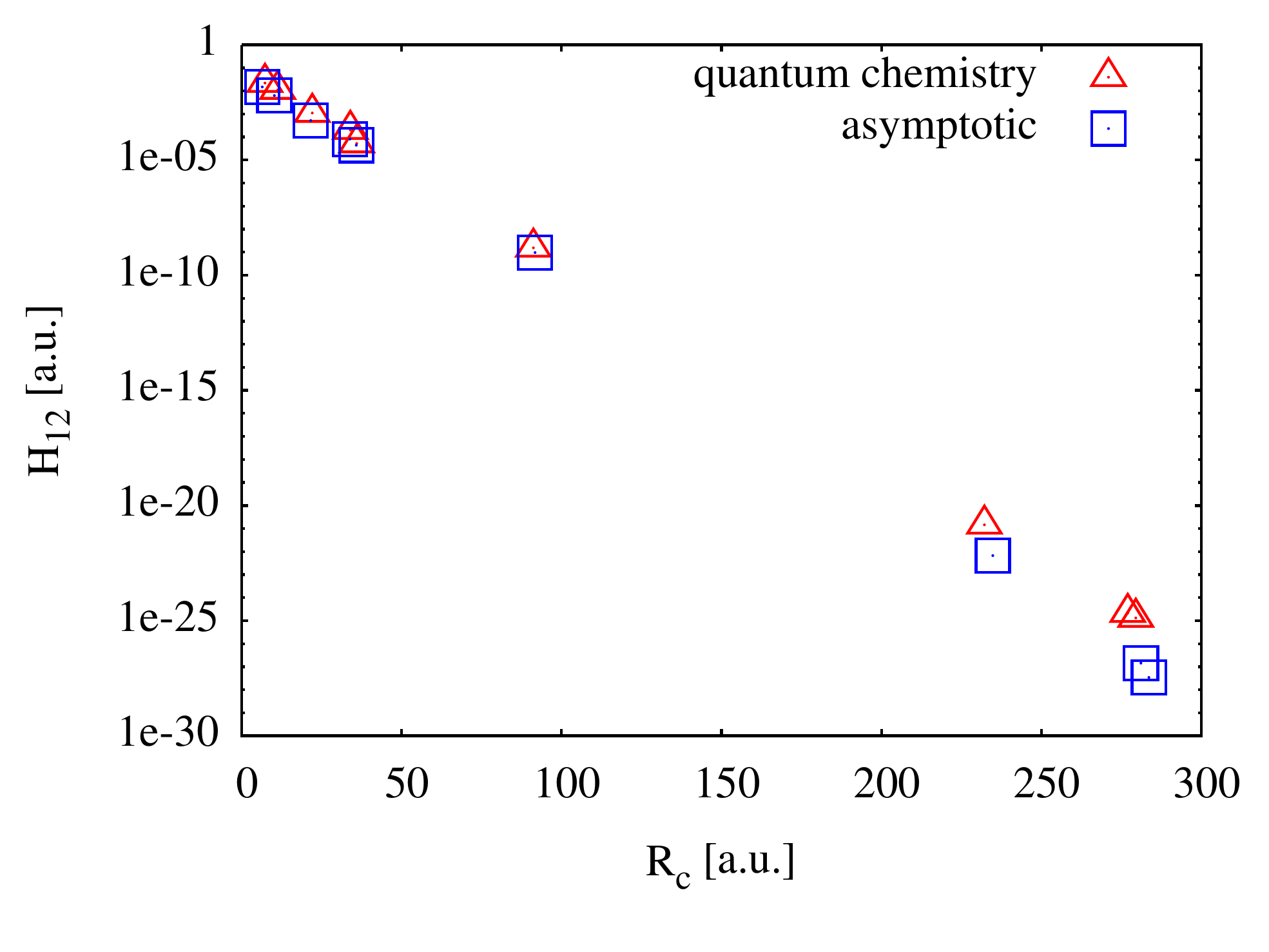}
\includegraphics[width=0.32\textwidth]{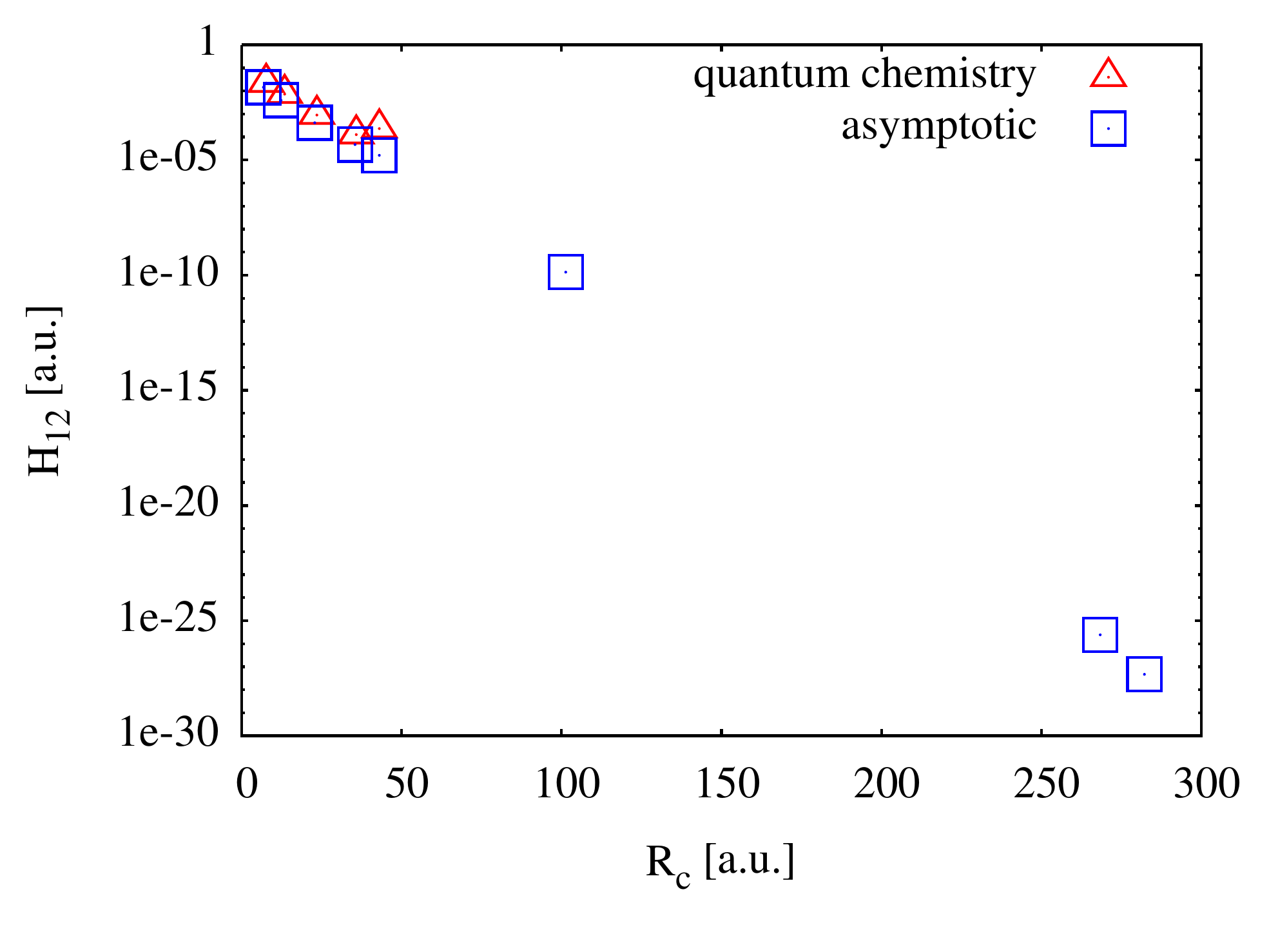}
\includegraphics[width=0.32\textwidth]{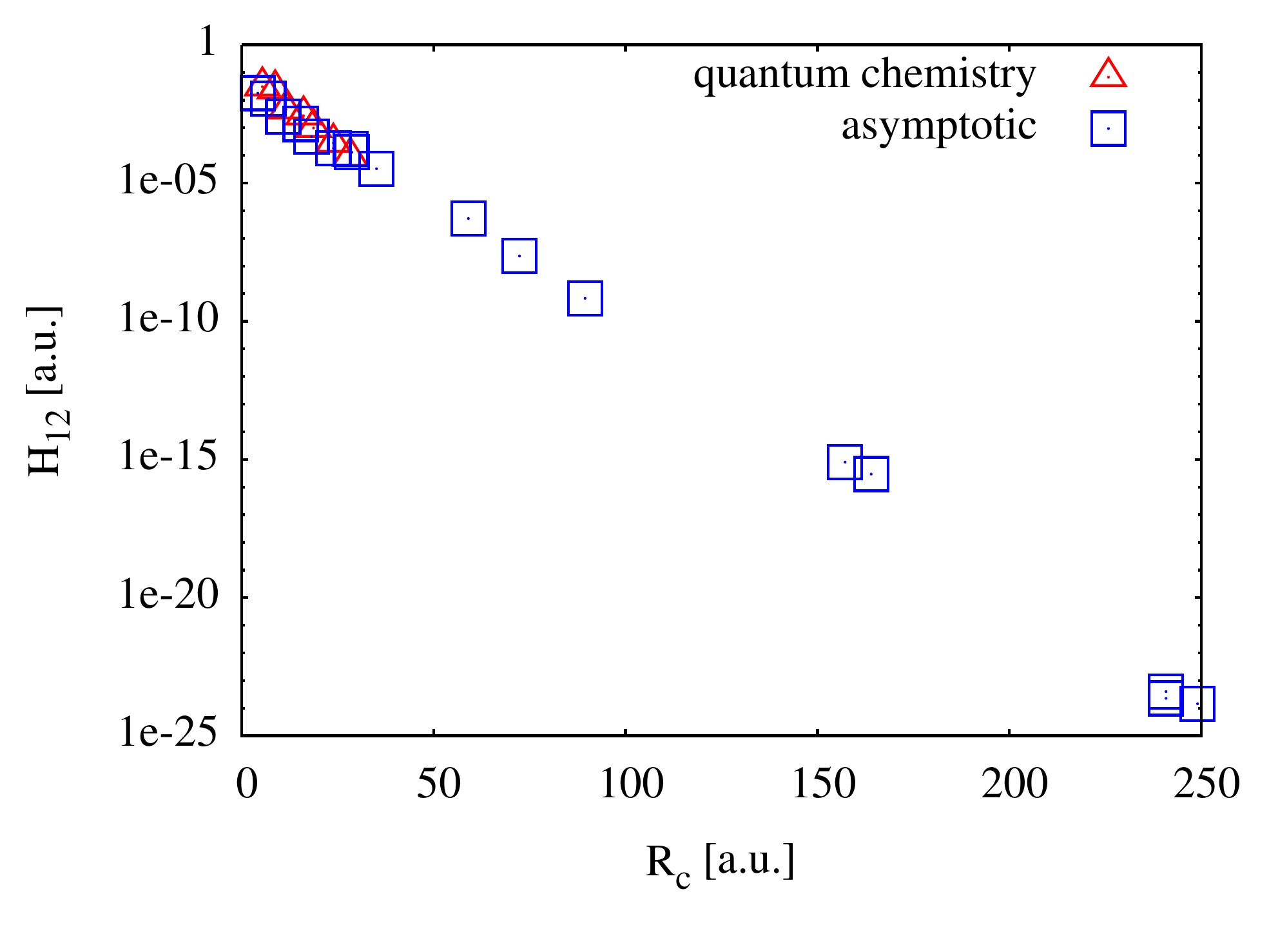}
\includegraphics[width=0.32\textwidth]{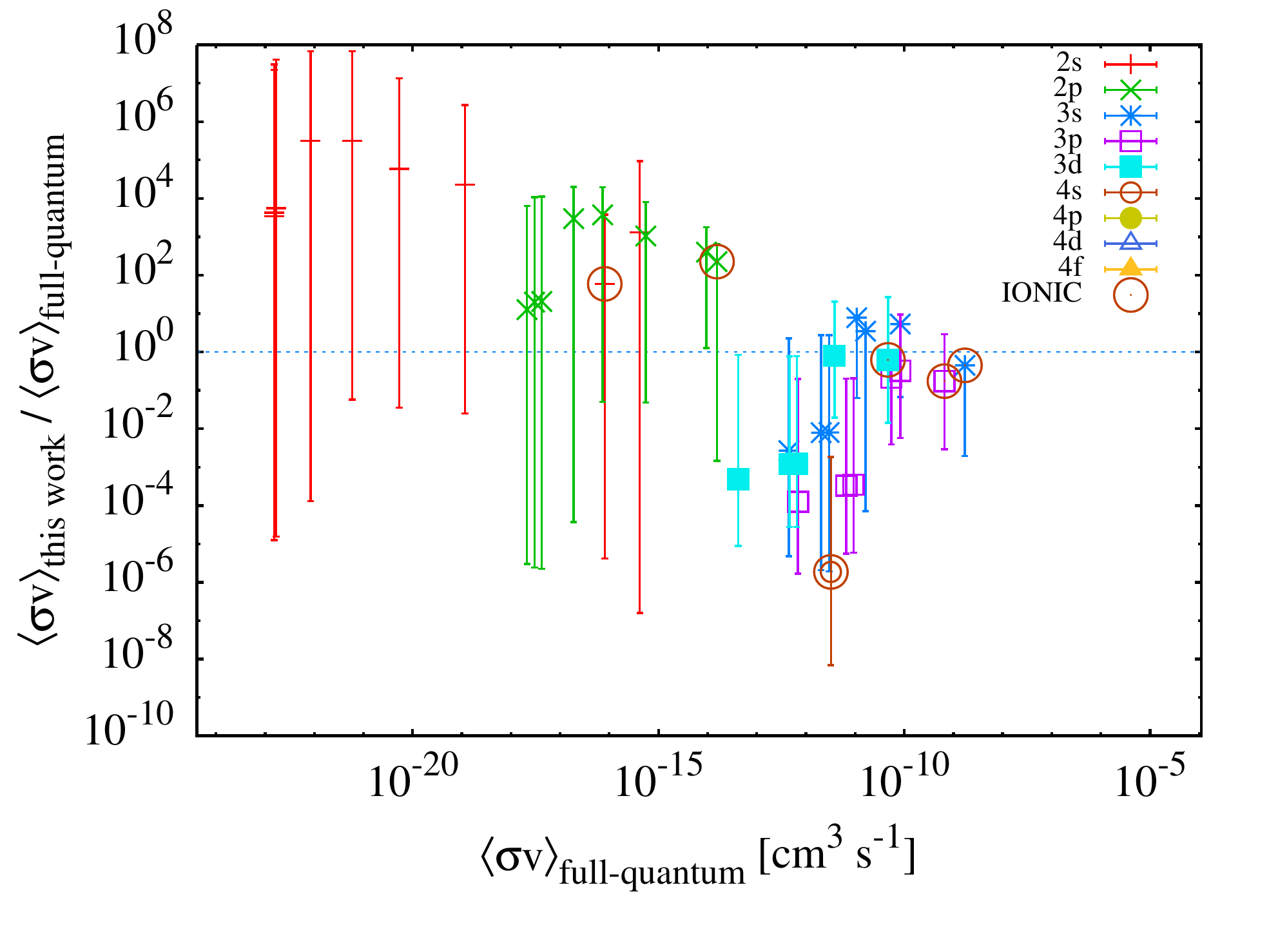}
\includegraphics[width=0.32\textwidth]{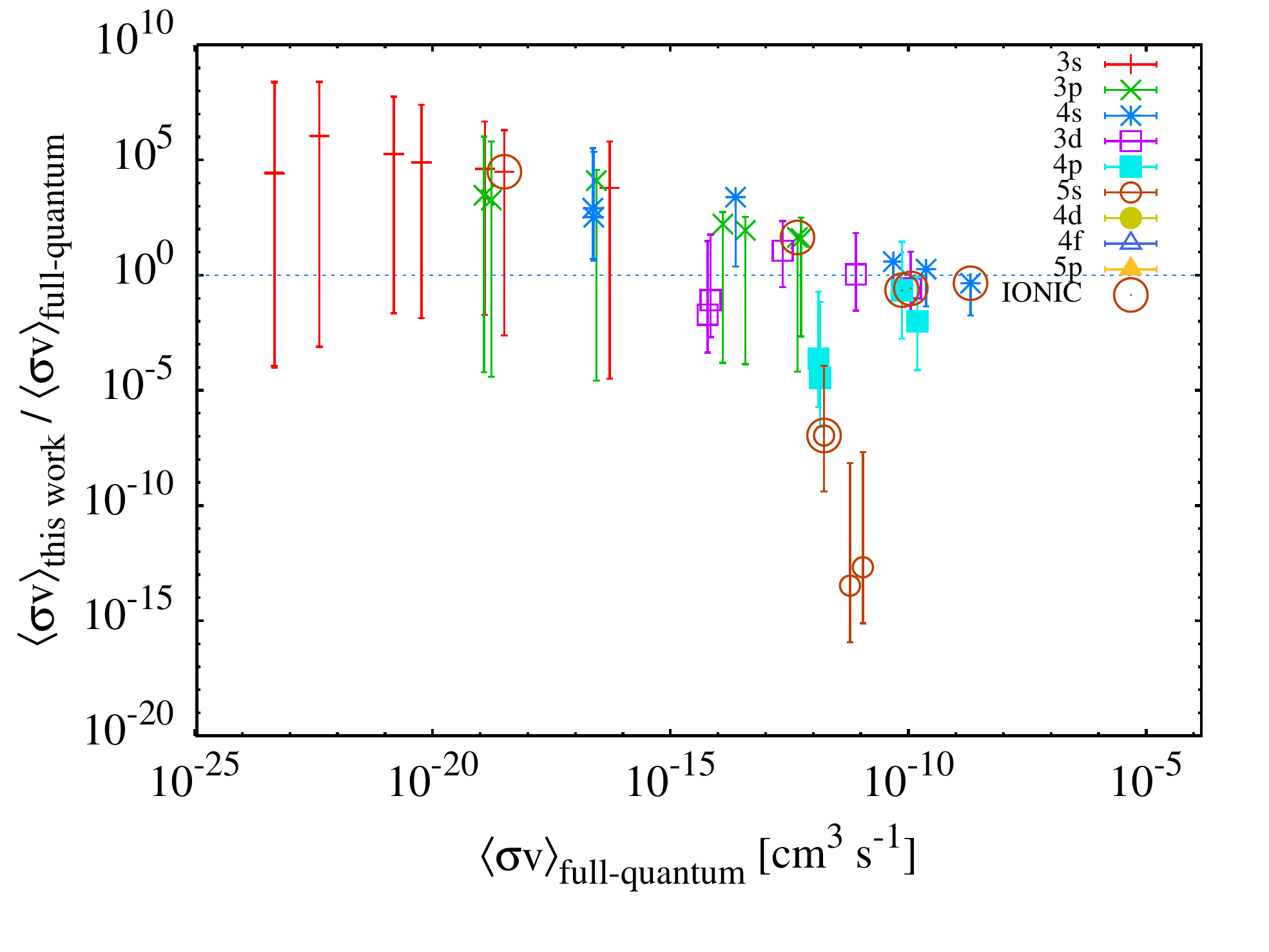}
\includegraphics[width=0.32\textwidth]{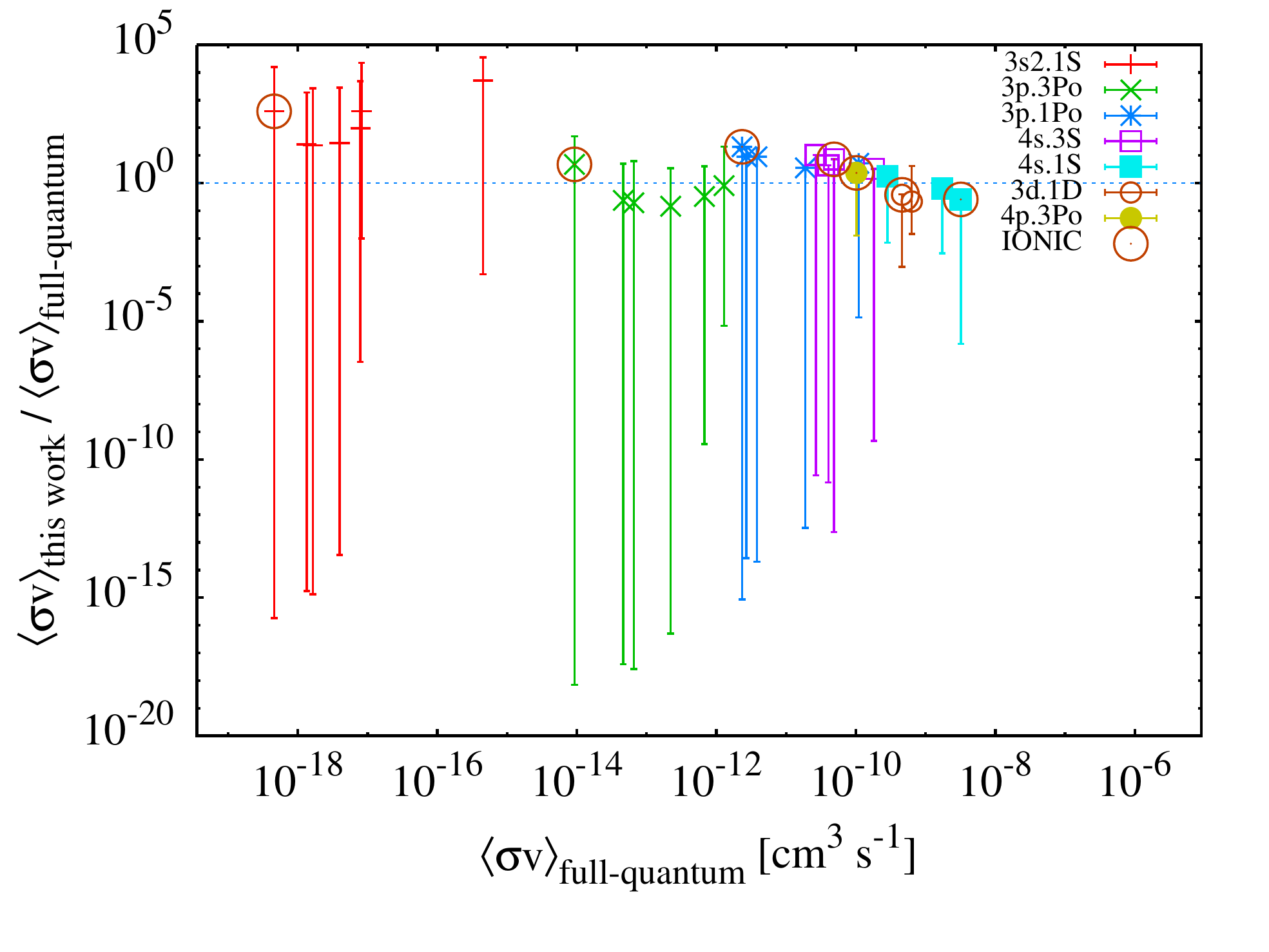}
\caption{Comparisons of calculations in the LCAO asymptotic model for Li+H, Na+H and Mg+H with results from full-quantum scattering calculations and the quantum chemistry data they are based on.  The first, second and third columns are for the Li+H, Na+H and Mg+H systems respectively.  The top row shows the adiabatic potentials for the LiH ($^1\Sigma^+$), NaH ($^1\Sigma^+$) and MgH ($^2\Sigma^+$) systems; the quantum chemistry potentials are from \cite{Belyaev2003,Croft1999a} for LiH, from \cite{dickinson_initio_1999} for NaH, and from \cite{Guitou2011} for MgH.  The middle row shows the coupling $H_{12}$; in the quantum chemistry calculations these are derived from the adiabatic potentials via the two-state Landau-Zener model (except for the long-range crossing data for Li+H, which are extrapolations of the data from shorter range, see \citep{Belyaev2003}).  The bottom row shows the ratio of rate coefficients at 6000~K, as a function of the full-quantum data; the full-quantum scattering results are from \cite{Belyaev2003,Barklem2003b} for Li+H, from \cite{Belyaev2010,Barklem2010} for Na+H, and from \cite{Belyaev2012,Barklem2012} for Mg+H.  The error bars show the ``fluctuations'', analogous to the uncertainty, calculated from the variation of alternate calculations as discussed in the text.}\label{fig:comp}
\end{figure*}

To test the asymptotic method presented here, the results of calculations for Li+H, Na+H, and Mg+H calculations are compared with these three sets of full-quantum calculation results.  Fig.~\ref{fig:comp} presents comparisons at three stages in the calculations: i) the potentials, ii) the LZ parameters, and iii) the final rate coefficients.  First, it is seen that the LCAO asymptotic method reproduces the relevant features of the interactions potentials at long range, i.e. the avoided crossings, quite well (and even some of the general behaviour at shorter range).  Second, the most important derived LZ parameters, the crossing point $R_c$ and the coupling at the crossing point $H_{12}$, are rather well reproduced compared to those derived from the quantum chemistry potentials.  There are some discrepancies for crossings at rather long range in Li+H; however, it should be noted that these are only extrapolations of the data from shorter range, see \citep{Belyaev2003}.   The crossing at $\sim 50$~a.u.\ in Na+H also shows some discrepancy, but this is most likely a deficiency in the resolution in $R$ of the quantum chemistry calculations, making it difficult to precisely calculate the LZ parameters from the potentials.  It is a significant advantage of the asymptotic approach that it can be easily calculated with any required resolution in $R$, up to the numerical precision of the codes.  Finally, the rate coefficients compare rather well, at least for the most important processes with large rate coefficients.  We note that for some processes the discrepancies are rather large, e.g. overestimated by a factor of $10^6$ for processes involving the 2$s$ state of Li. However, these processes have extremely small rate coefficients, of order $10^{-20}$~cm$^3$ s$^{-1}$, and even if they were a factor of $10^6$ larger, they would still be insignificant compared to the dominant processes with rate coefficients of order $10^{-8}$~cm$^3$ s$^{-1}$.

In order to get an objective, quantitative measure of the success of the LCAO aymptotic model and the other model approaches to estimate the rate coefficients (Semi-emp, LH-S, LH-J), two statistical goodness-of-fit measures are calculated comparing to the existing full-quantum scattering calculations.  The first is the usual $\chi^2$ statistic, calculated by:
\beq
\chi^2 = \sum (\la \sigma v \ra_\mathrm{model} - \la \sigma v \ra_\mathrm{full-quantum})^2 ,
\eeq
where the summation is over all transitions for which there is data in the full-quantum calculations.
The second accounts for the fact that large rates are likely to be more important in applications, and each term is weighted by the full-quantum rate coefficient:
\beq
\chi^2_w = \sum \la \sigma v \ra_\mathrm{full-quantum} (\la \sigma v \ra_\mathrm{model} - \la \sigma v \ra_\mathrm{full-quantum})^2 .
\eeq
The results for these two goodness-of-fit measures are given for all four models in Table~\ref{tab:chisq}. Clearly the LCAO model performs best in all three cases, irrespective of whether $\chi^2$ or $\chi_w^2$ is used, and thus is our preferred model.  Generally, the Semi-emp model performs next best, followed by LH-J, and finally LH-S. In \cite{MiyanoHedberg2014} Semi-emp, LH-J and LH-S model results were compared with data obtained \emph{ab initio} quantum mechanically on $\mathrm{H}^+ + \mathrm{H}^−$, and concluded that the LH-J expressions gave the most reliable results.   We note that comparison of the results of the preferred model LCAO, with the other three models provides interesting information on the sensitivity of the results to modelling.  For all calculations, we calculate the results with all four models, as well as the LCAO model using the final adopted LZ parameters as well as the diabatic and adiabatic ones, and then extract the maximum and minimum obtained values for a given process.  In the lower panels of Fig.~\ref{fig:comp}, these maximum and minimum values are plotted as error bars on the LCAO results.  It is not to be implied that these give an accurate measure of the uncertainty in the calculations, but these results do allow an estimate of how much rate coefficients might vary, similar to the``fluctuation factor'' presented in \cite{Barklem2010} and which is somewhat analogous to an estimate of the uncertainty.   It is seen that in almost all cases in Fig.~\ref{fig:comp}, these minimum and maximum values, which we will call ``fluctuations'', bound the full-quantum result, and perhaps overestimate the uncertainty.

\begin{table*}
\caption{\label{tab:chisq} Goodness-of-fit measures $\chi^2$ and $\chi^2_w$ for the four models compared to the existing full-quantum calculations, for hydrogen collision processes with Li, Na and Mg.  }
\begin{ruledtabular}
\begin{tabular}{lcccccc}
 & \multicolumn{2}{c}{Li+H} & \multicolumn{2}{c}{Na+H} & \multicolumn{2}{c}{Mg+H} \\
              Model & $\chi^2$ & $\chi^2_w$ & $\chi^2$ & $\chi^2_w$ & $\chi^2$ & $\chi^2_w$ \\ \hline
              &&&&&&\\
            LCAO    & $2.96\times 10^{-15}$ & $1.35\times 10^{-29}$ & $2.46\times 10^{-15}$ & $1.34\times 10^{-29}$ & $3.02\times 10^{-15}$ & $4.83\times 10^{-30}$ \\
          Semi-emp  & $7.77\times 10^{-15}$ & $3.72\times 10^{-29}$ & $7.05\times 10^{-15}$ & $4.10\times 10^{-29}$ & $3.53\times 10^{-15}$ & $8.25\times 10^{-30}$ \\
            LH-S    & $1.92\times 10^{-14}$ & $4.26\times 10^{-29}$ & $2.24\times 10^{-14}$ & $4.37\times 10^{-29}$ & $3.48\times 10^{-15}$ & $9.14\times 10^{-30}$ \\
            LH-J    & $1.63\times 10^{-14}$ & $3.96\times 10^{-29}$ & $1.25\times 10^{-14}$ & $2.47\times 10^{-29}$ & $3.47\times 10^{-15}$ & $9.14\times 10^{-30}$ \\
\end{tabular}
\end{ruledtabular}
\end{table*}

\begin{figure*}
\centering
\includegraphics[width=0.32\textwidth]{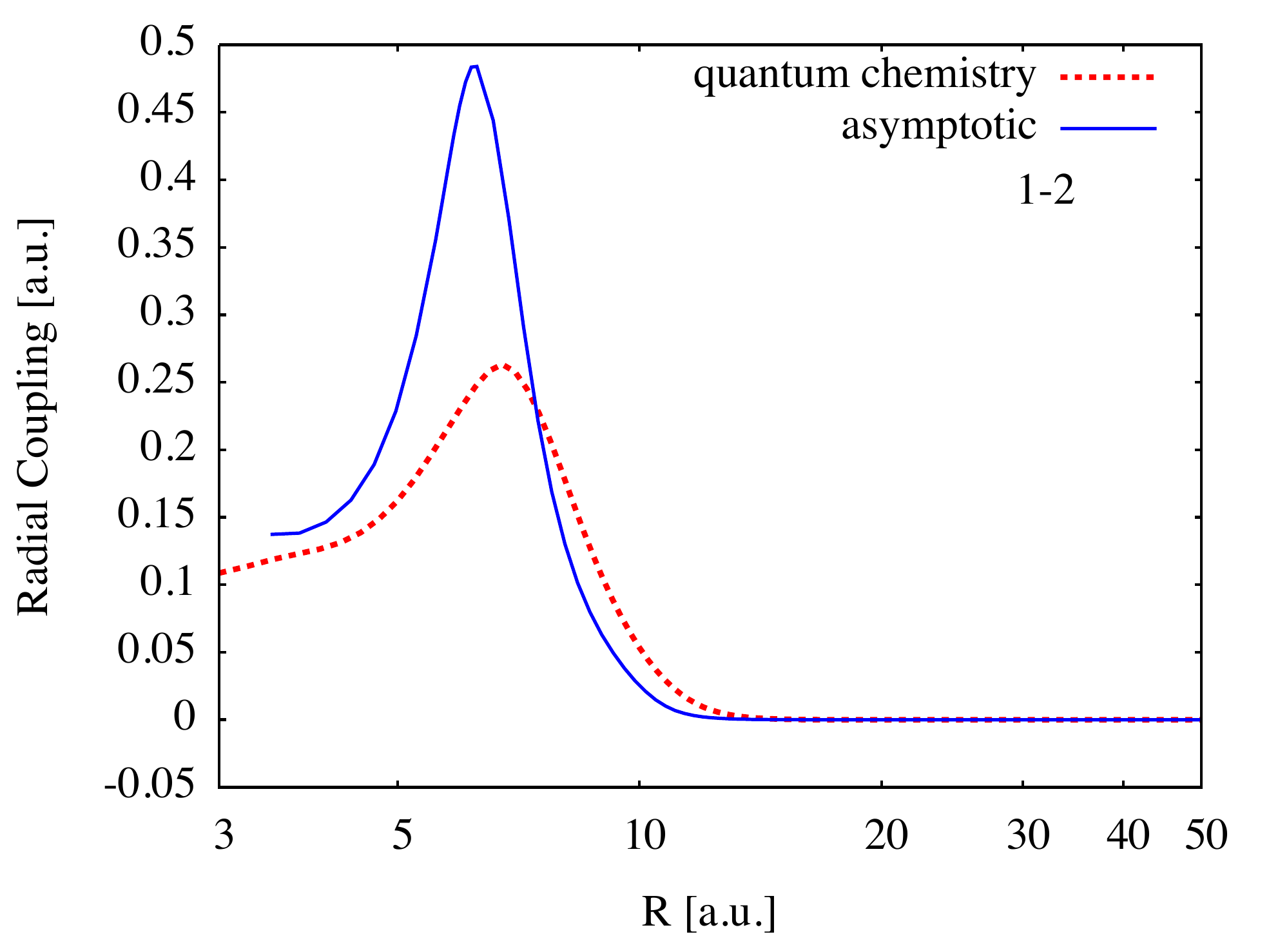}
\includegraphics[width=0.32\textwidth]{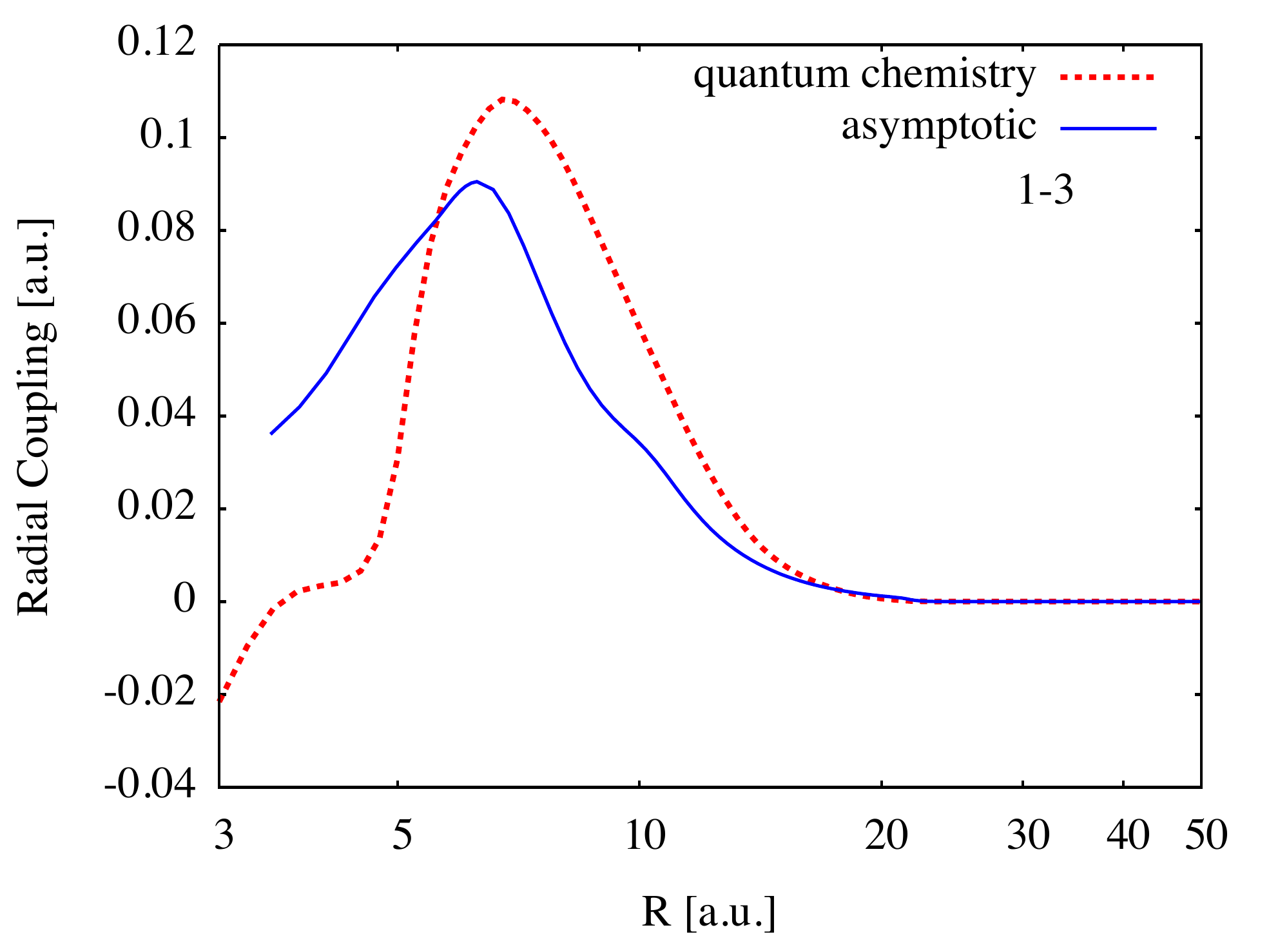}
\includegraphics[width=0.32\textwidth]{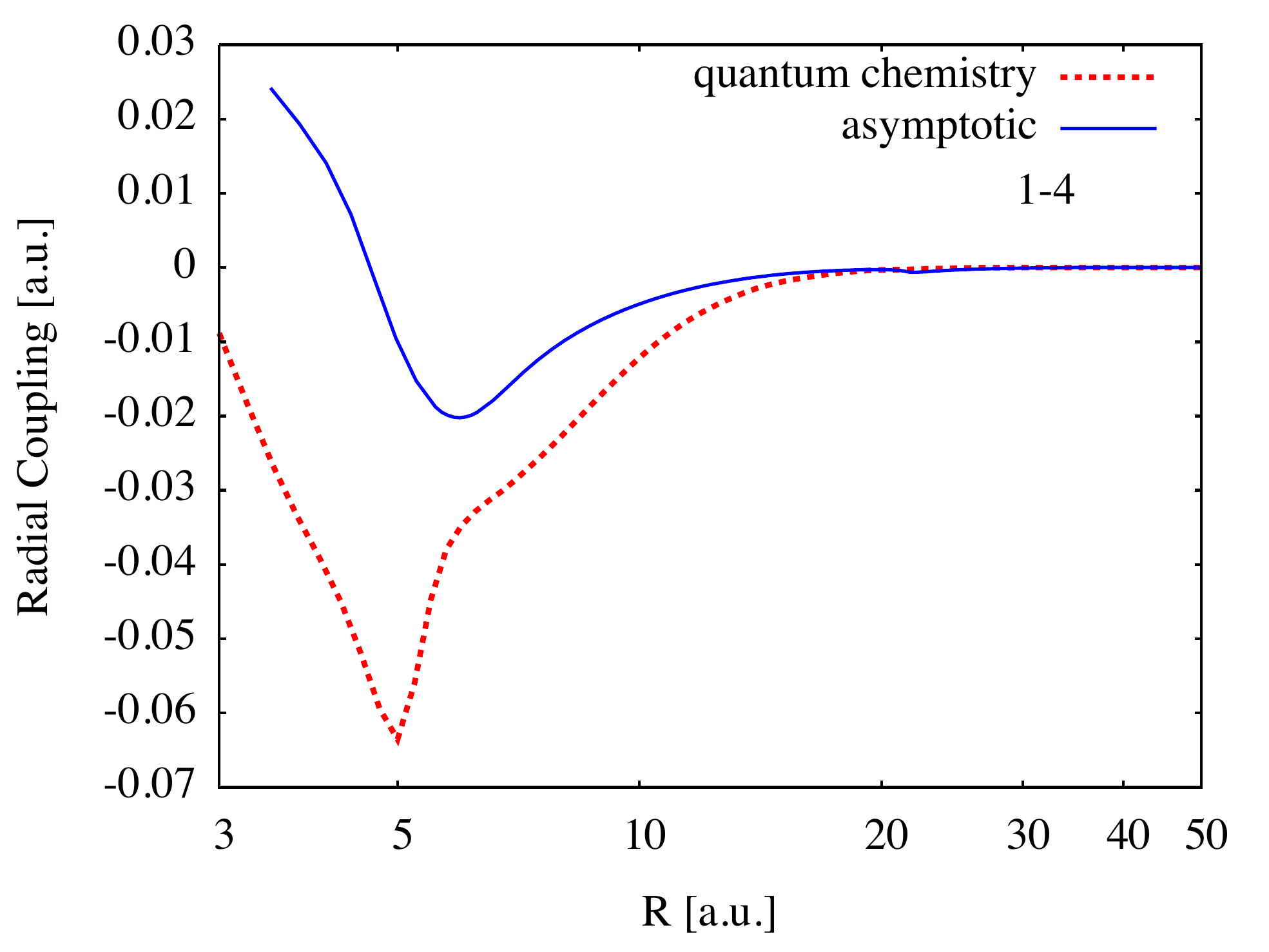}
\includegraphics[width=0.32\textwidth]{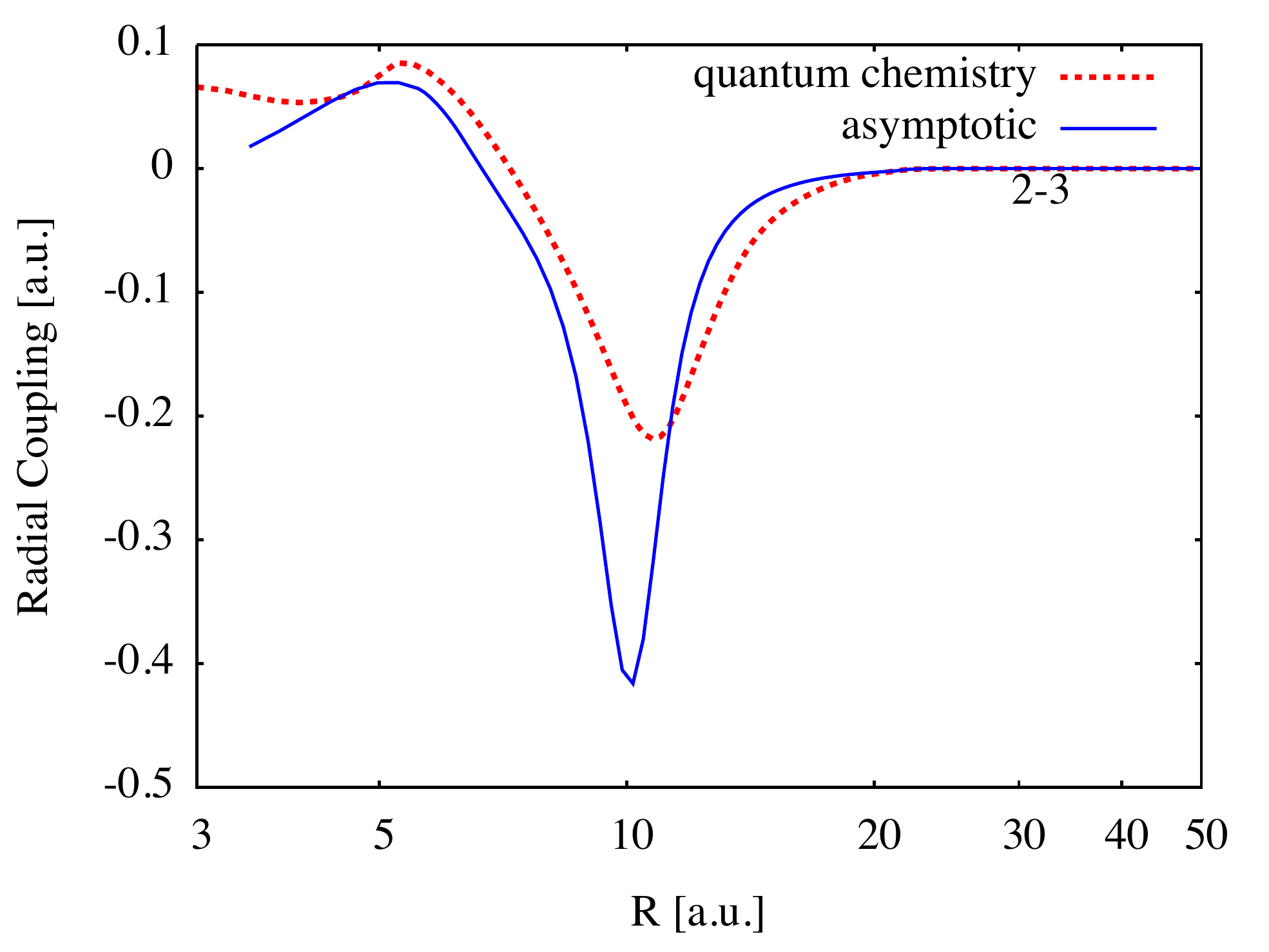}
\includegraphics[width=0.32\textwidth]{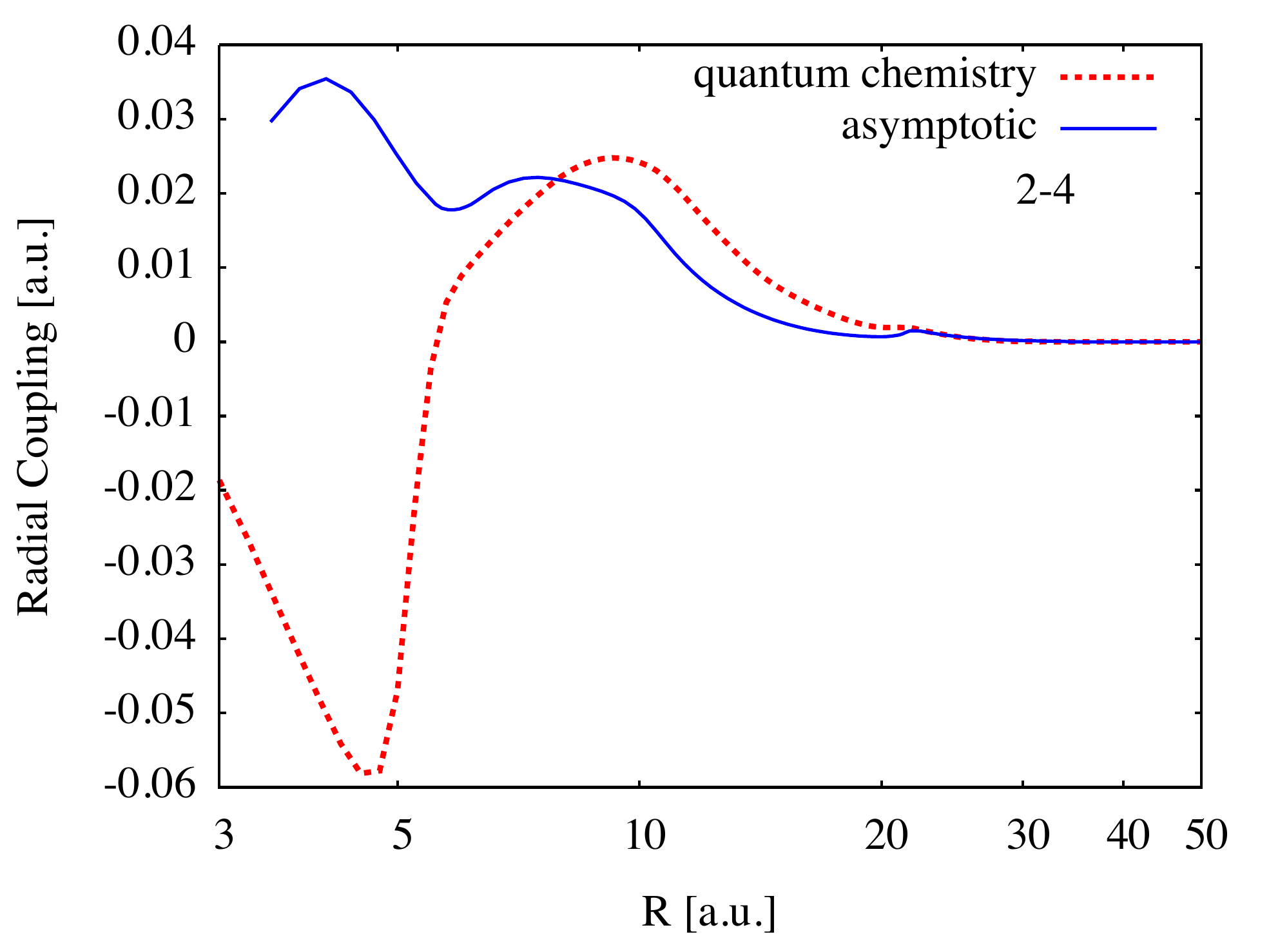}
\includegraphics[width=0.32\textwidth]{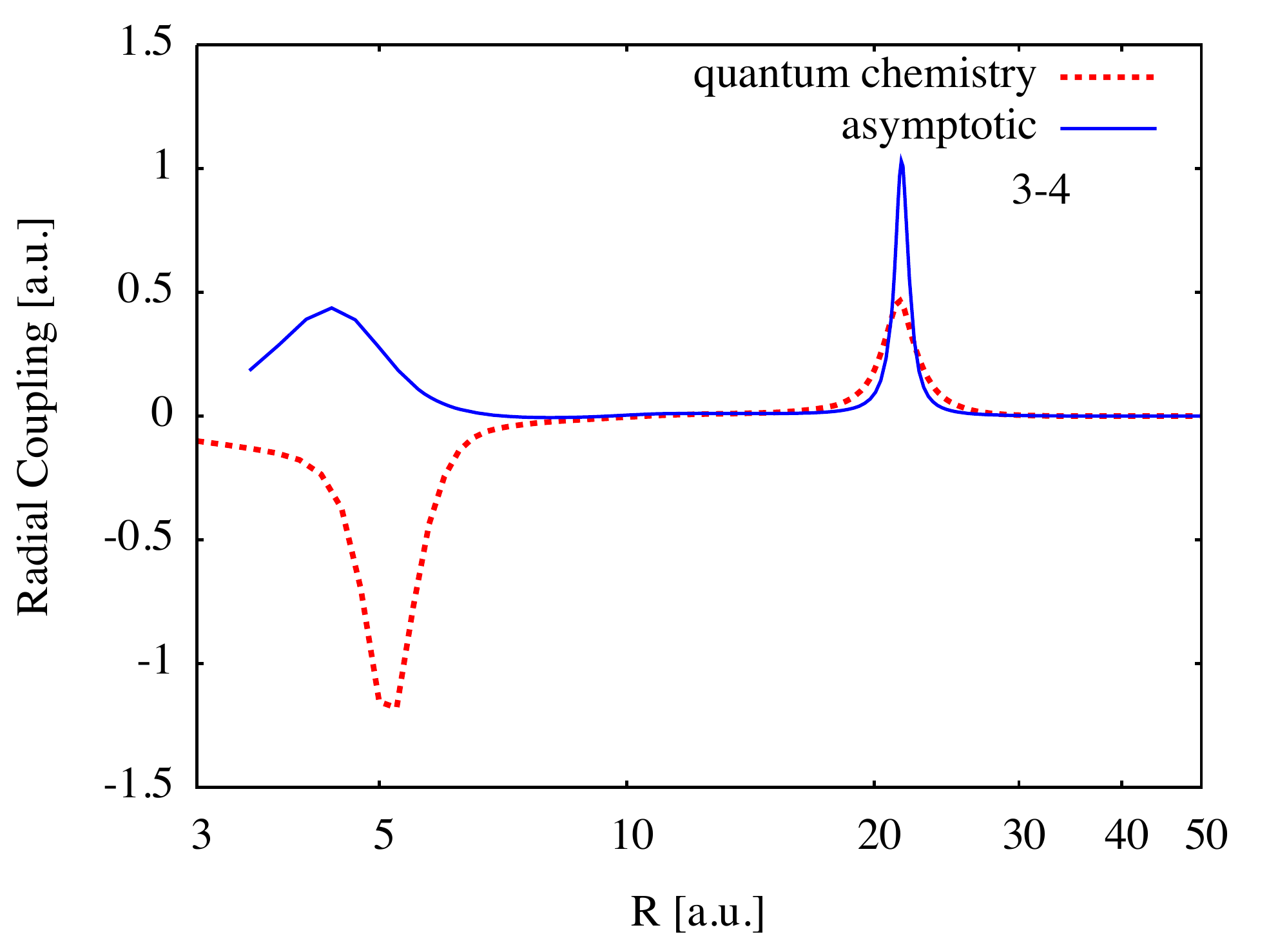}
\caption{Comparison of radial couplings for the LiH($^1\Sigma^+$) quasimolecule.  The nonadiabatic radial coupling matrix elements between the states X, A, C, D $^1\Sigma^+$ (labelled 1,2,3,4 here) are compared with data from \cite{Belyaev2003} based on quantum chemistry calculations of \cite{Croft1999a}, which are improvements on earlier work   \citep{Gadea1993,boutalib_abinitio_1992}. }\label{fig:ddrLi}
\end{figure*}

Finally, though not used in our dynamical calculations, it is of interest to compare the radial coupling results.  As mentioned, radial coupling results from the asymptotic model could conceivably be used in a hybrid approach combined with quantum chemistry results for low-lying states.  Fig.~\ref{fig:ddrLi} compares the results for Li+H for some couplings between low-lying states, with the quantum chemistry results from \cite{Belyaev2003} based on quantum chemistry calculations of \cite{Croft1999a}, which are improvements on earlier work \citep{Gadea1993,boutalib_abinitio_1992}.  The results are encouraging with the forms, magnitudes, and widths of the couplings at the ionic crossings reproduced to within roughly a factor of two.  Comparisons for Na+H and Mg+H are similar. 

\subsection{Calculations for Ca}
\label{sect:Ca_res}

\begin{figure*}
\centering
\begin{overpic}[width=0.32\textwidth]{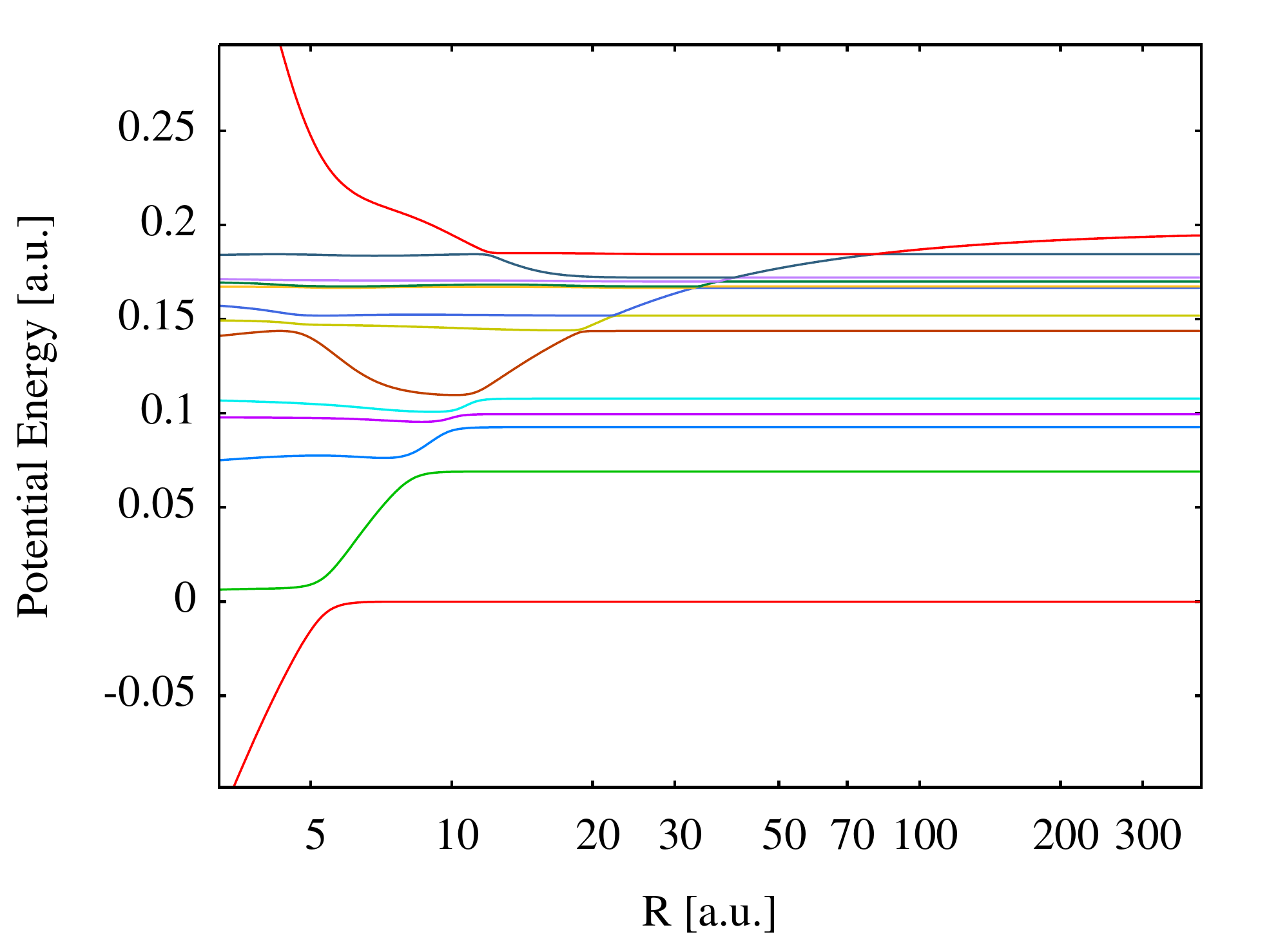}\put(25,17){$^2\Sigma^+$, $\mathrm{Ca}^+(4s)$ core}\end{overpic}
\begin{overpic}[width=0.32\textwidth]{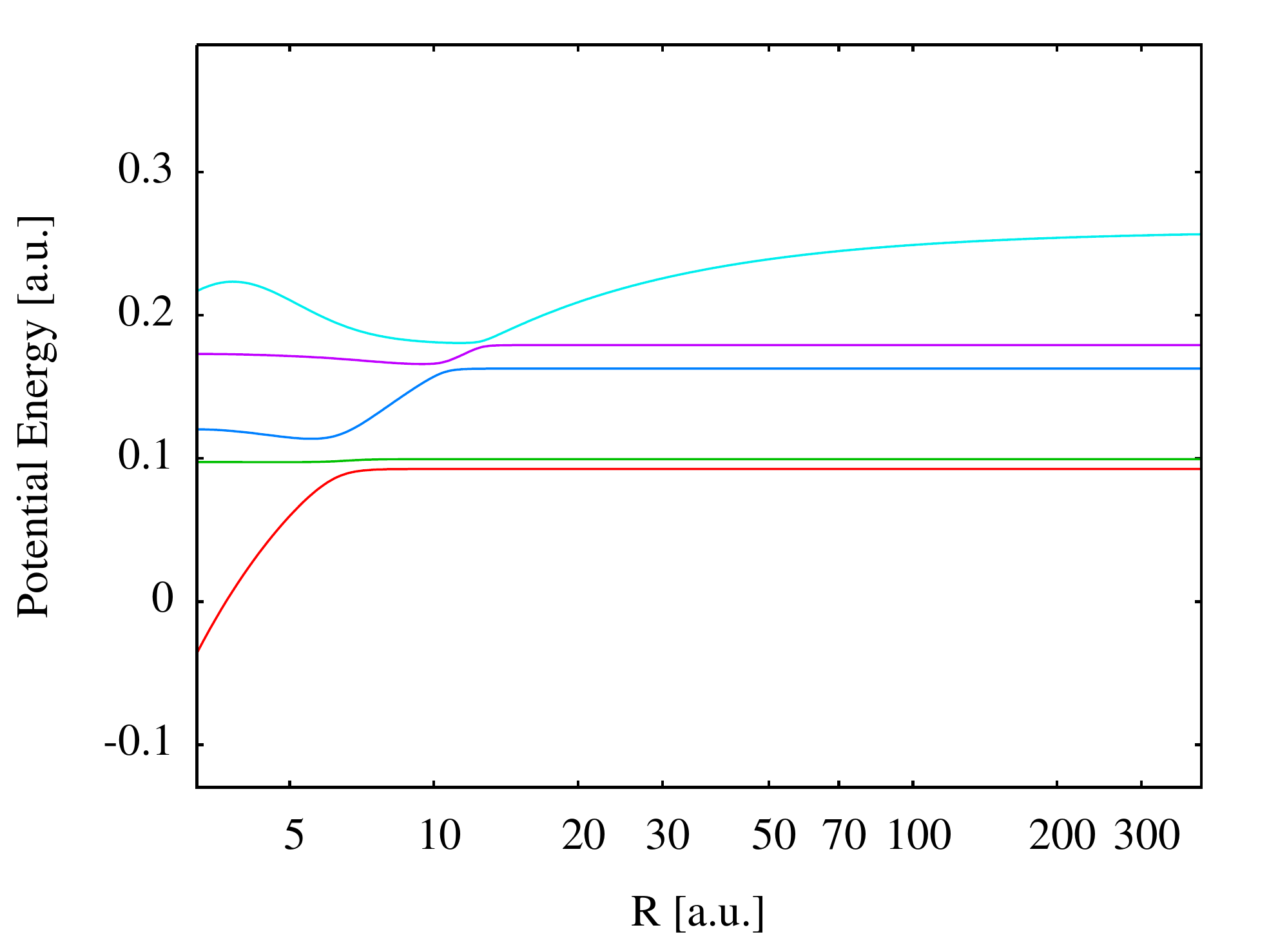}\put(25,17){$^2\Sigma^+$, $\mathrm{Ca}^+(3d)$ core}\end{overpic}
\begin{overpic}[width=0.32\textwidth]{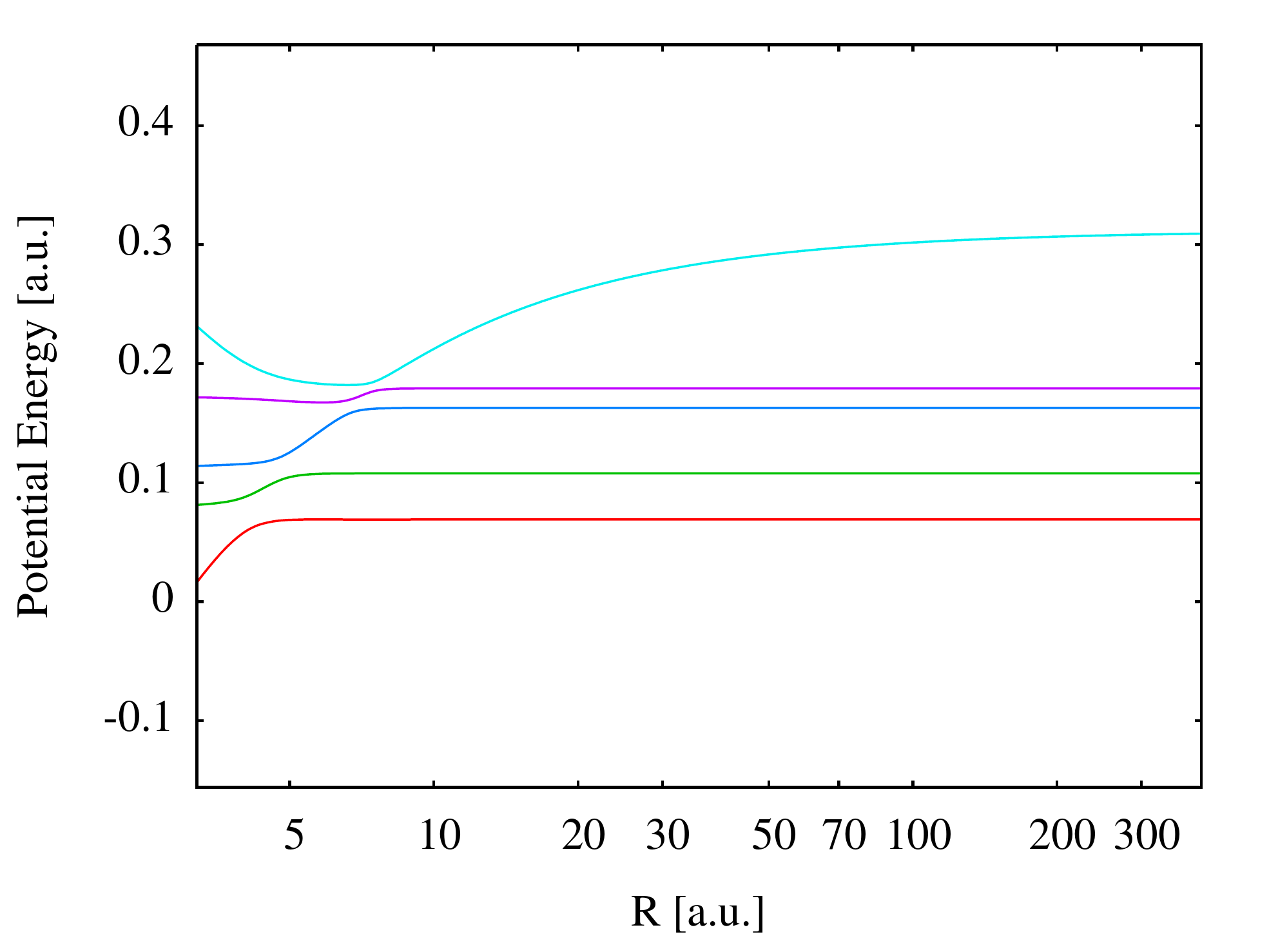}\put(25,17){$^2\Sigma^+$, $\mathrm{Ca}^+(4p)$ core}\end{overpic}
\begin{overpic}[width=0.32\textwidth]{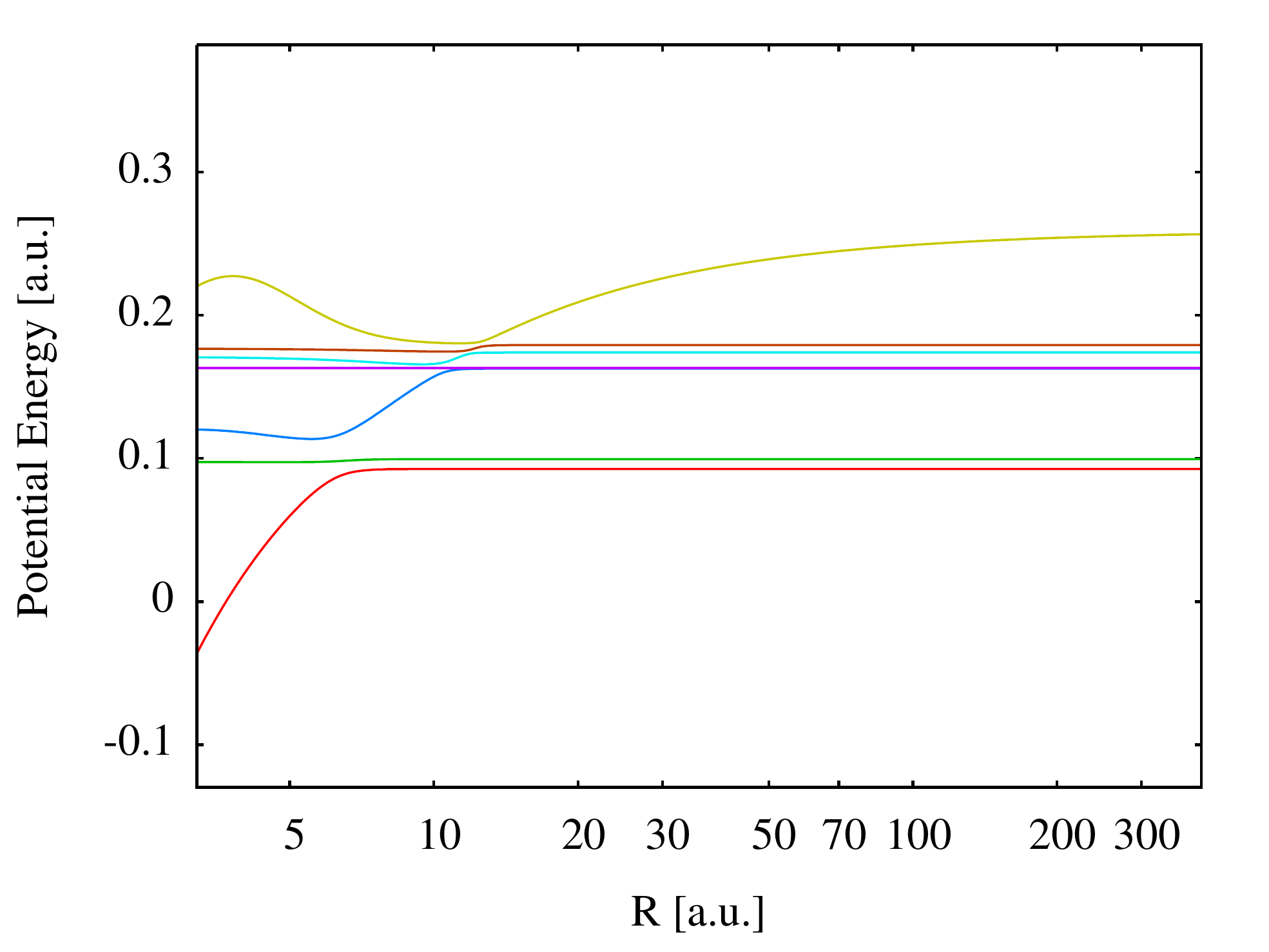}\put(25,17){$^2\Pi$, $\mathrm{Ca}^+(3d)$ core}\end{overpic}
\begin{overpic}[width=0.32\textwidth]{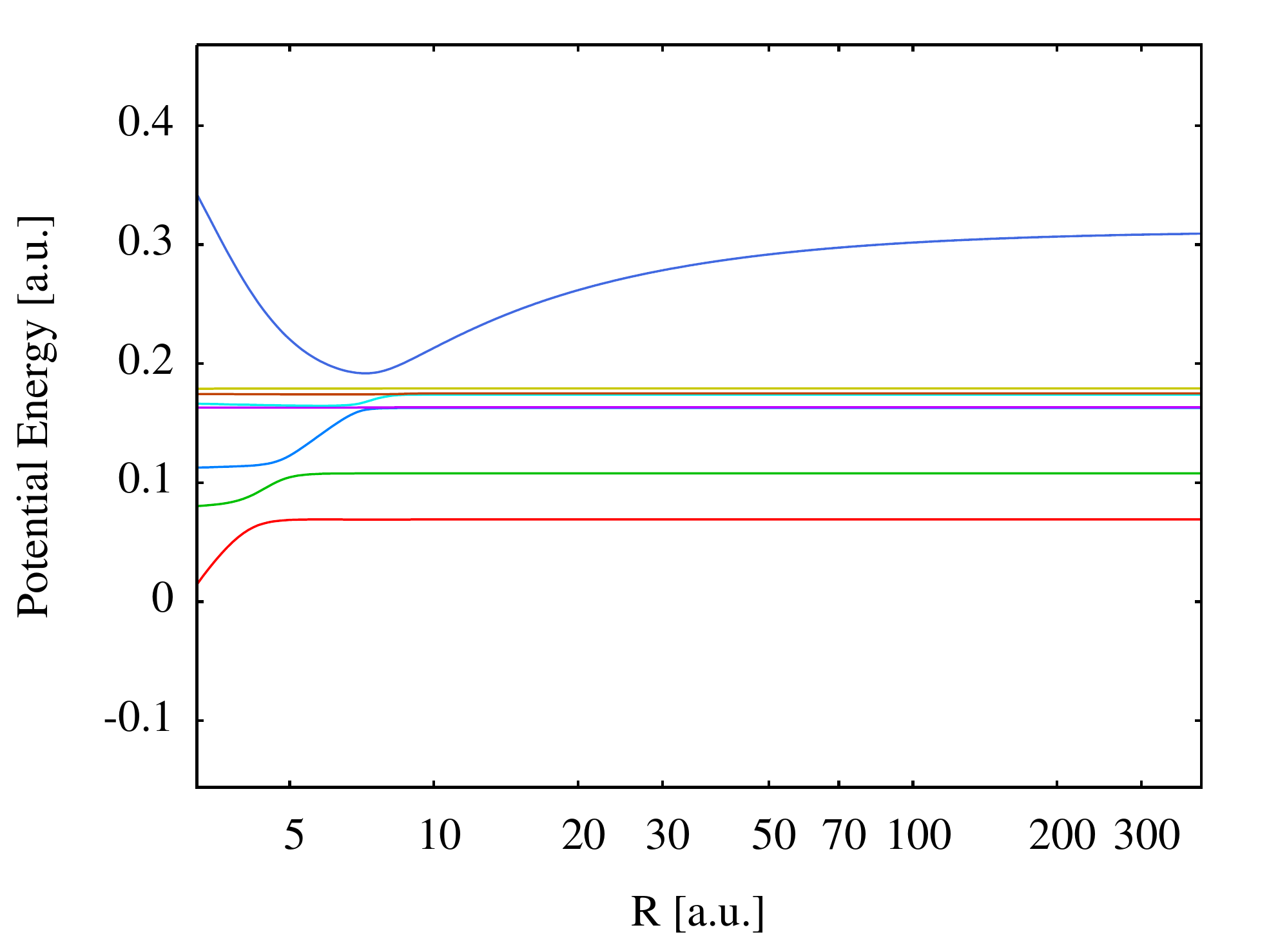}\put(25,17){$^2\Pi$, $\mathrm{Ca}^+(4p)$ core}\end{overpic}
\begin{overpic}[width=0.32\textwidth]{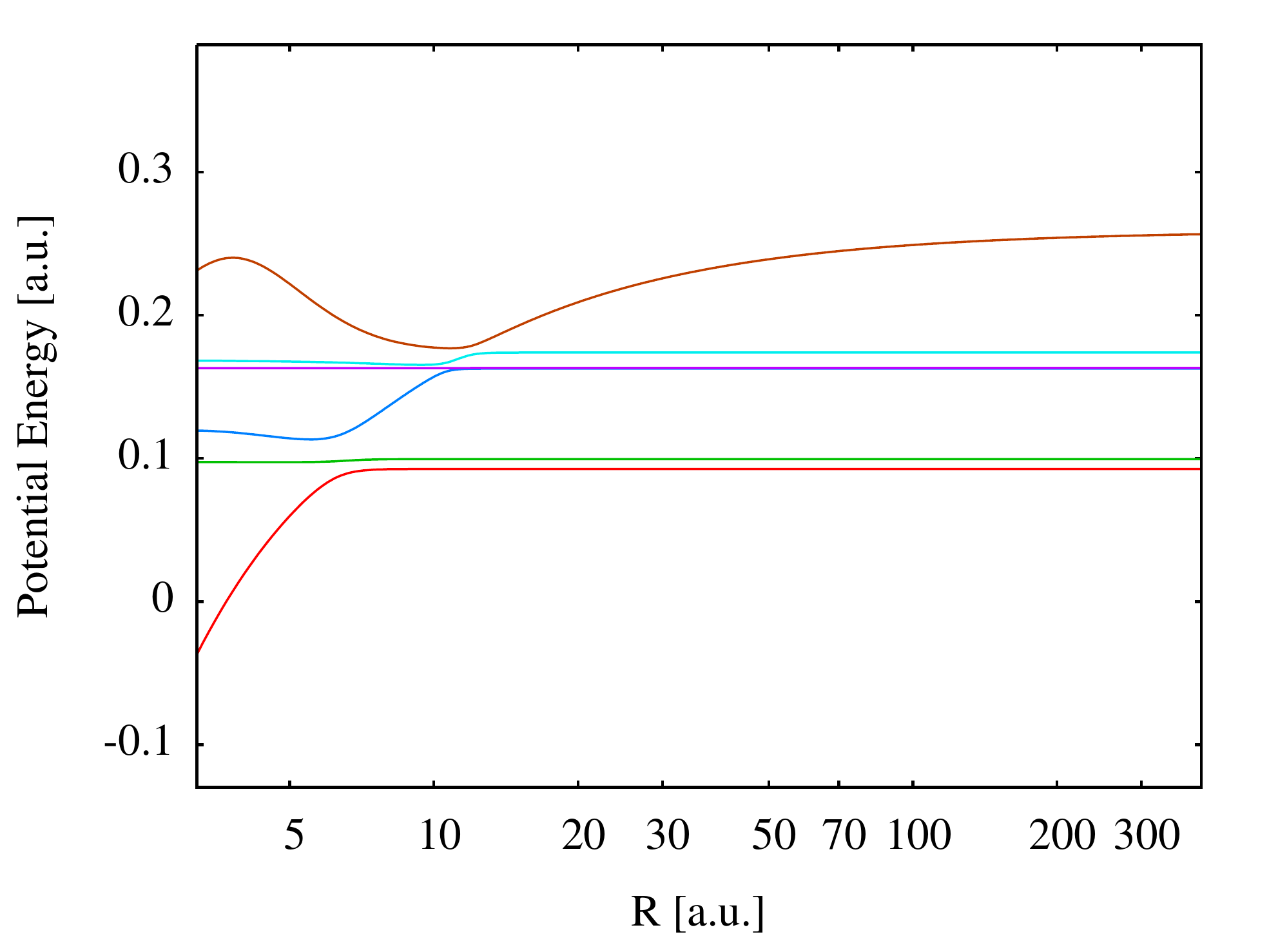}\put(25,17){$^2\Delta$, $\mathrm{Ca}^+(3d)$ core}\end{overpic}
\caption{The potentials energies for Ca+H.  The top row is $^2\Sigma^+$, with cores $\mathrm{Ca}^+(4s)$, $\mathrm{Ca}^+(3d)$, and $\mathrm{Ca}^+(4p)$.  The bottom row are $^2\Pi$, with cores $\mathrm{Ca}^+(3d)$, $\mathrm{Ca}^+(4p)$, and $^2\Delta$, with core $\mathrm{Ca}^+(3d)$.  }\label{fig:Ca_pots}
\end{figure*}

\begin{figure*}
\centering
\begin{overpic}[width=0.32\textwidth]{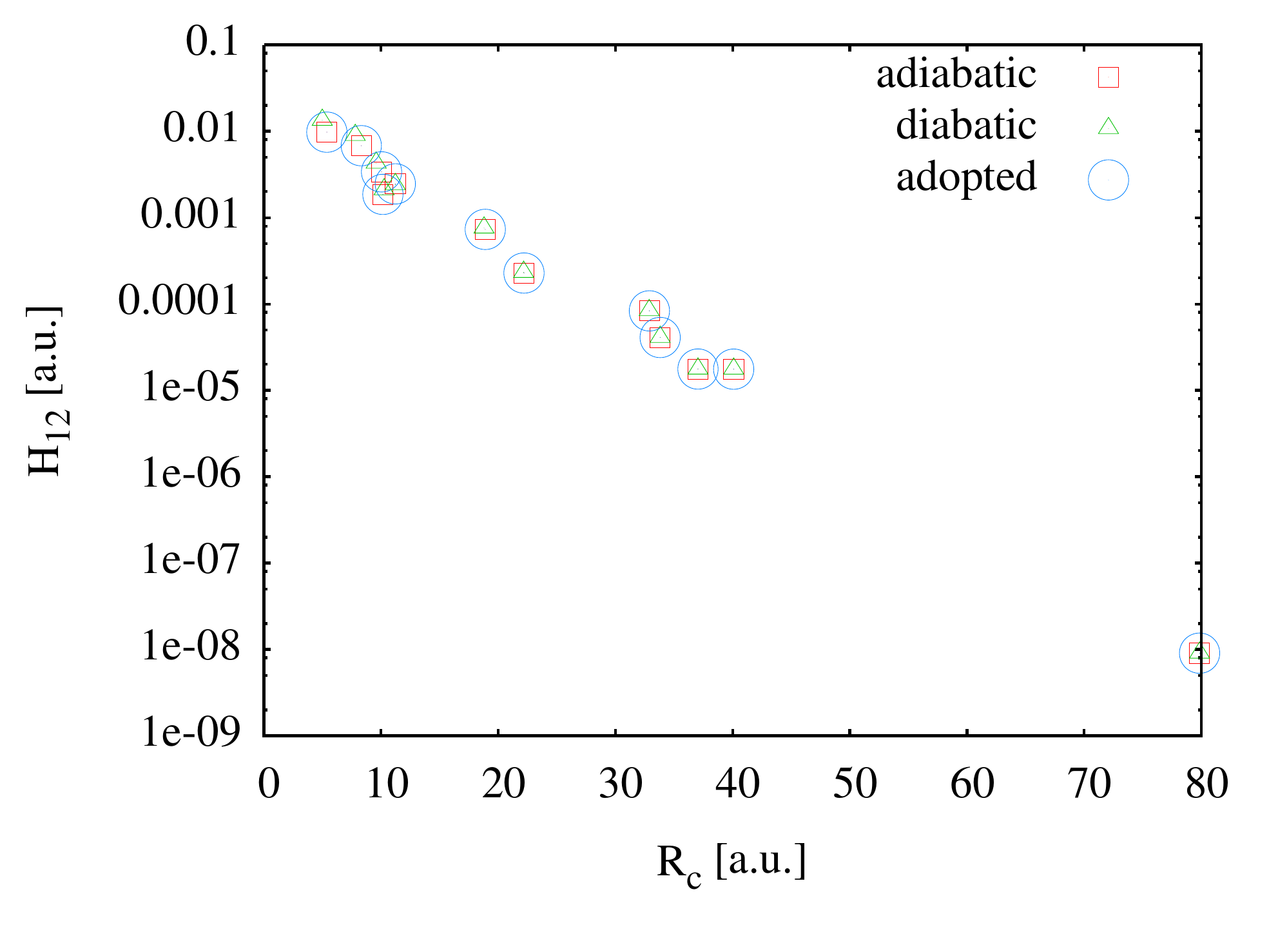}\put(25,20){$^2\Sigma^+$, $\mathrm{Ca}^+(4s)$ core}\end{overpic}
\begin{overpic}[width=0.32\textwidth]{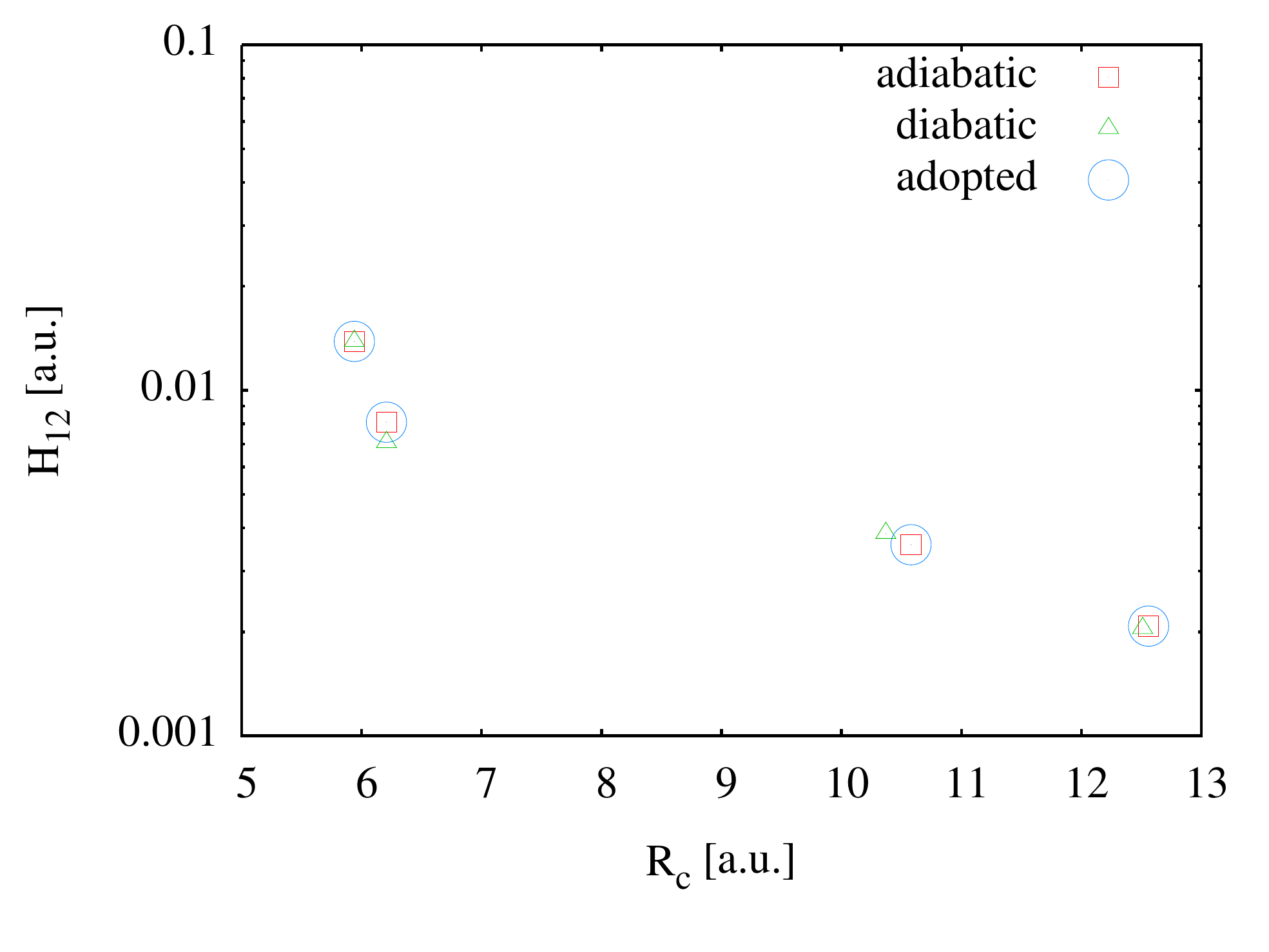}\put(25,20){$^2\Sigma^+$, $\mathrm{Ca}^+(3d)$ core}\end{overpic}
\begin{overpic}[width=0.32\textwidth]{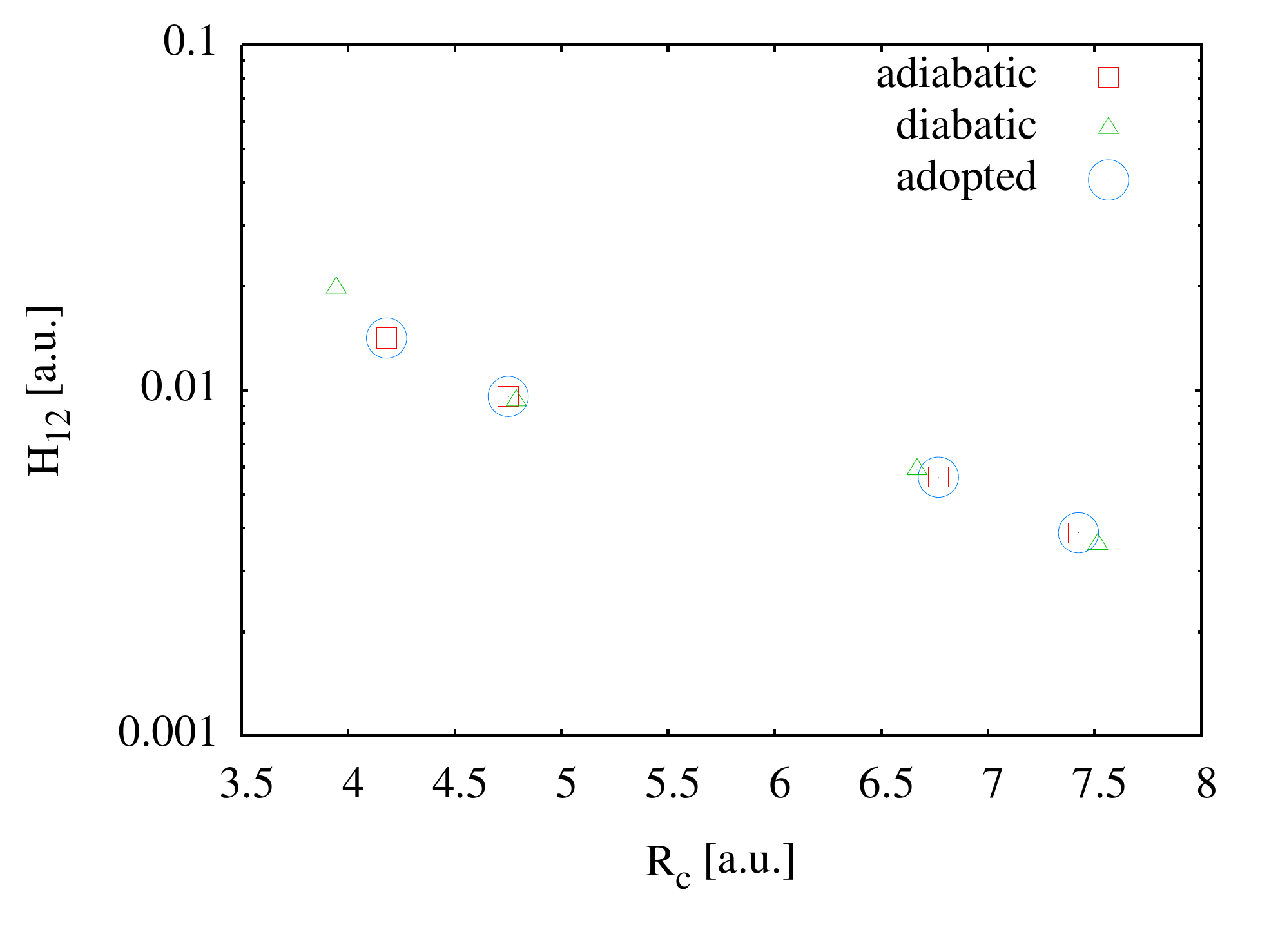}\put(25,20){$^2\Sigma^+$, $\mathrm{Ca}^+(4p)$ core}\end{overpic}
\begin{overpic}[width=0.32\textwidth]{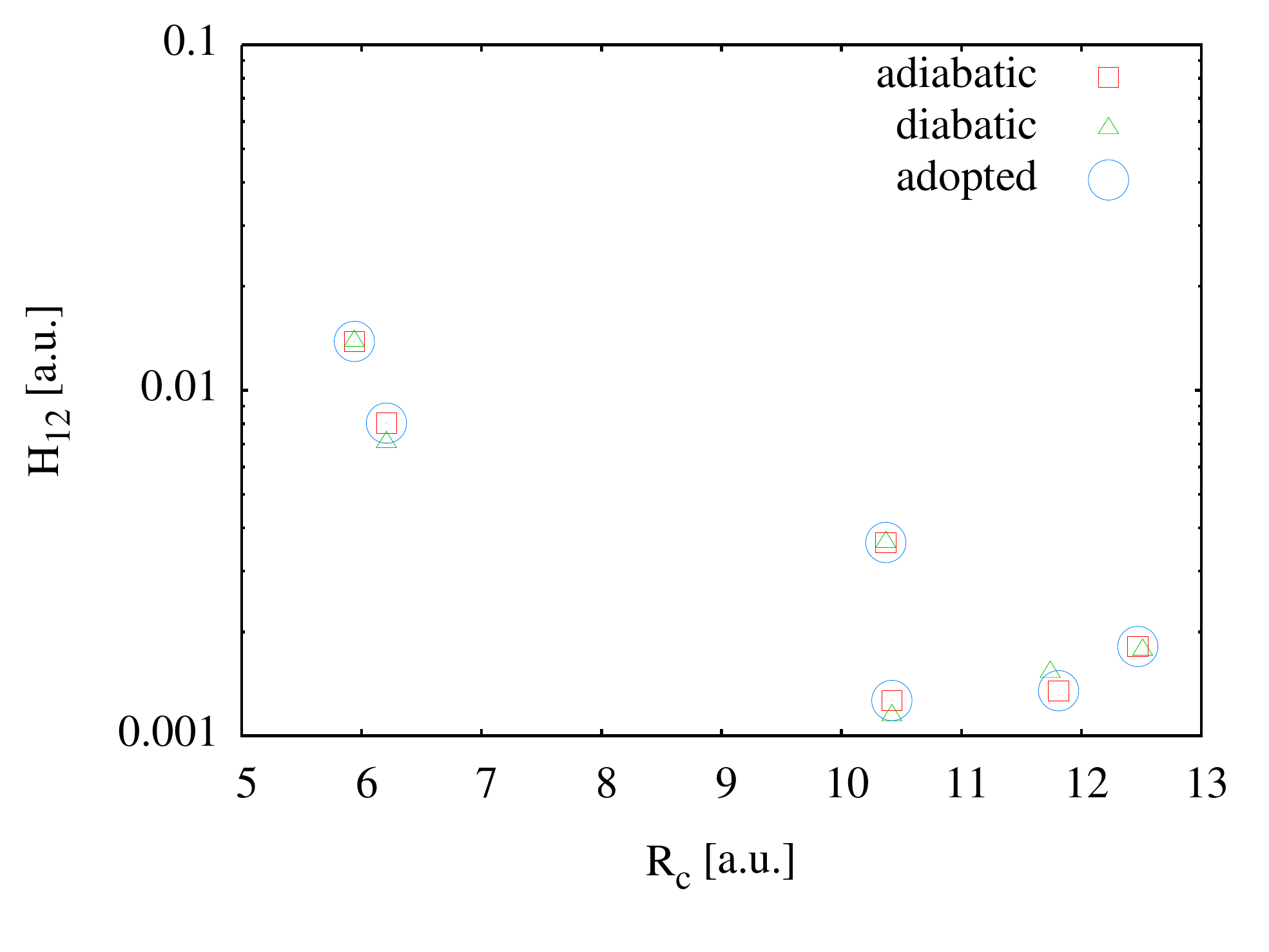}\put(25,20){$^2\Pi$, $\mathrm{Ca}^+(3d)$ core}\end{overpic}
\begin{overpic}[width=0.32\textwidth]{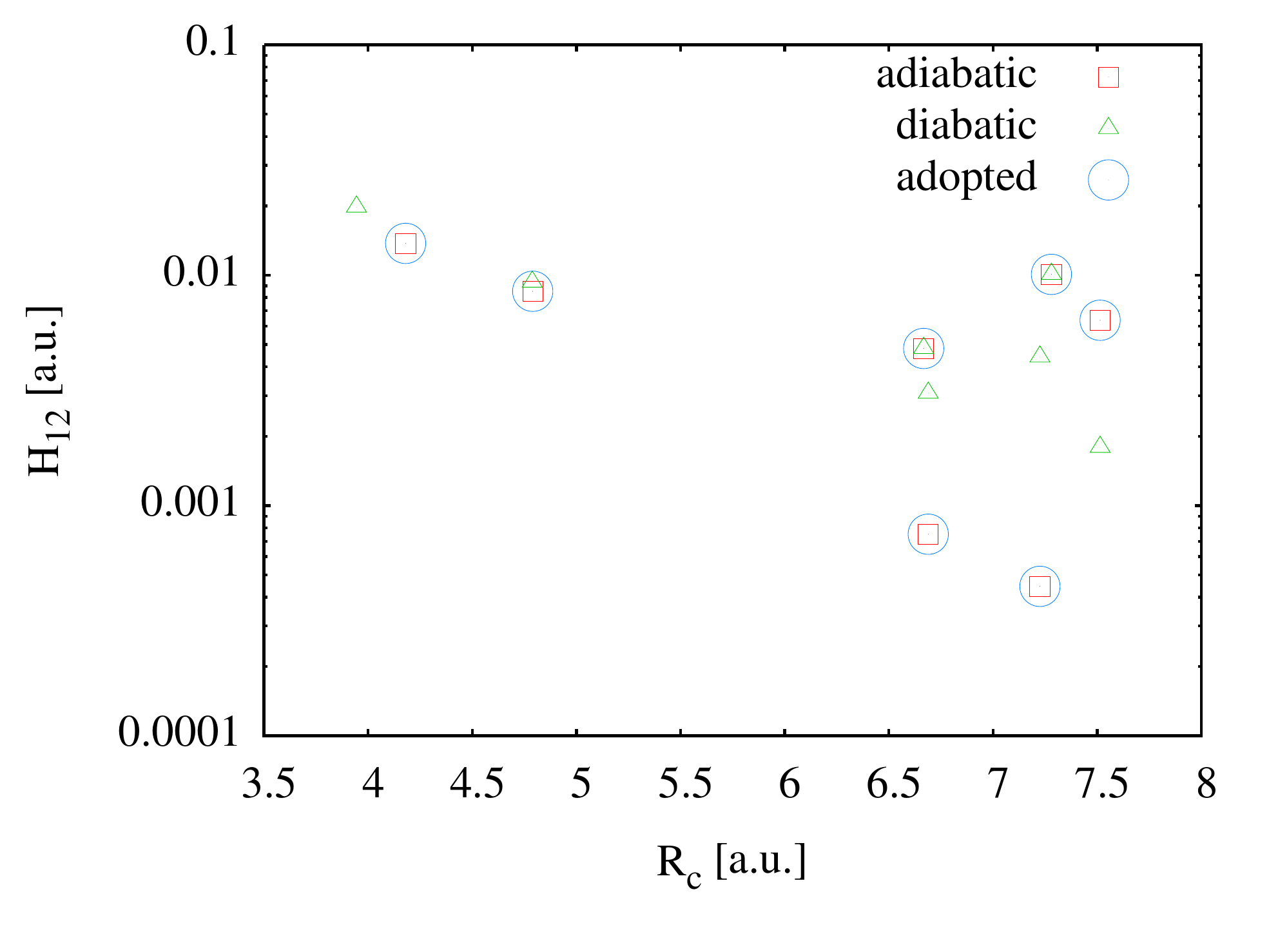}\put(25,20){$^2\Pi$, $\mathrm{Ca}^+(4p)$ core}\end{overpic}
\begin{overpic}[width=0.32\textwidth]{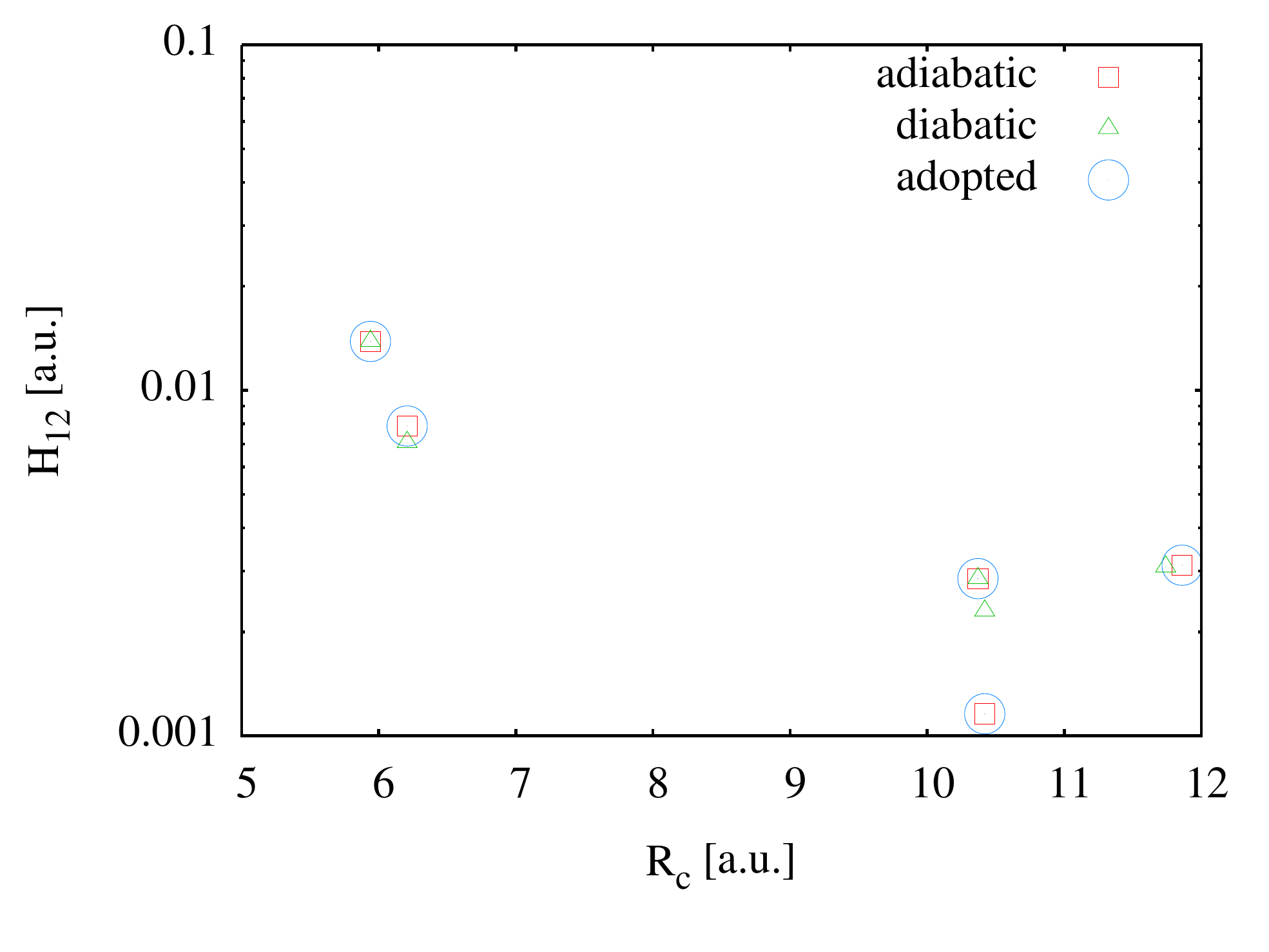}\put(25,20){$^2\Delta$, $\mathrm{Ca}^+(3d)$ core}\end{overpic}
\caption{The LZ parameters for Ca+H.  The top row is $^2\Sigma^+$, with cores $\mathrm{Ca}^+(4s)$, $\mathrm{Ca}^+(3d)$, and $\mathrm{Ca}^+(4p)$.  The bottom row are $^2\Pi$, with cores $\mathrm{Ca}^+(3d)$, $\mathrm{Ca}^+(4p)$, and $^2\Delta$, with core $\mathrm{Ca}^+(3d)$.  }\label{fig:Ca_lz}
\end{figure*}

Calcium is an element of significant astrophysical importance, and at this time there are no calculations of Ca+H inelastic processes suitable for non-LTE modelling of the Ca spectrum in cool stars \footnote{Just prior to submission, a paper on this subject appearred \cite{belyaev_model_2016}.  Comparison with the calculations presented here should be the subject of future work.}.  It is an obvious candidate for a first application of the method presented here.  In Table~\ref{tab:ca_input}, the input data for the calculations are presented, with the majority extracted from the NIST atomic spectra database \citep{kramida_current_2014,NIST,sugar_atomic_1985}.  The coefficients of fractional parentage are always unity in this case.  Three core states of Ca$^+$ are considered, as these can potentially lead to ionic crossings of Ca$^+$+H$^-$ configurations with covalent configurations of Ca+H corresponding to low-lying states of Ca at intermediate to large internuclear distance, and thus processes with significant cross sections.  More excited cores lead only to crossings at very short internuclear distance, and thus cannot lead to large cross sections.  Table~\ref{tab:ca_syms} lists the possible symmetries arising for this system, including the three that need to be calculated in the asymptotic model for the three considered cores, along with the relevant statistical weights.  The resulting potential energy curves for the six symmetry-core combinations that need to be calculated are shown in Fig.~\ref{fig:Ca_pots}, and the derived LZ parameters $R_c$ and $H_{12}$ in Fig.~\ref{fig:Ca_lz}.  Fig.~\ref{fig:Ca_lz} shows LZ parameters calculated both from the adiabatic and diabatic data, in addition to the adopted values.  The main differences occur in cases where states lie very close to each other, thus creating a series of very narrow crossings, such as in the case of $^2\Pi$ with core Ca$^+$(4p).  The differences are at worst an order of magnitude in $H_{12}$, which is significant.  In such cases, an argument could be made that the diabatic values are to be preferred. However, we note that the two-state LZ model is not strictly appropriate in such a case, and that a calculation using the LCAO diabatic LZ parameter data are among the calculations used to determine the fluctuations, and will therefore influence these values if the impact of assuming the adiabatic values would be large.

\begin{table*}
\caption{\label{tab:ca_input}  Input data for Ca+H calculations.  The zero point in the case of \ion{Ca}{i} states is its ground state, and the zero point for states involved in ionic configurations is the  \ion{Ca}{ii} ground state.  In the case of covalent configurations, the hydrogen atom ground state, H($1s$), is implied and for clarity not written. }
\begin{ruledtabular}
\begin{tabular}{lccccrrclccc}
$         \mathrm{Configuration}$ & $         L_A$ & $      2S_A+1$ & $           n$ & $           l$ & $           E$ & $     E_{lim}$ & $      N_{eq}$ & $          \mathrm{Core}$ & $         L_c$ & $      2S_c+1$ & $    G^{S_A L_A}_{S_c L_c}$ \\ 
  &  &  &  &  & [cm$^{-1}$] & [cm$^{-1}$] &  &  &  &  &  \\ \hline\
$                      4s^2.^1S$ & $           0$ & $           1$ & $           4$ & $           0$ & $           0$ & $       49305$ & $           2$ & $                       \mathrm{Ca}^+.4s$  &$           0$ & $           2$ & $                    1.000$ \\
$                    4s4p.^3P^o$ & $           1$ & $           3$ & $           4$ & $           1$ & $       15158$ & $       49305$ & $           1$ & $                       \mathrm{Ca}^+.4s$  &$           0$ & $           2$ & $                    1.000$ \\
$                    4p4s.^3P^o$ & $           1$ & $           3$ & $           4$ & $           0$ & $       15158$ & $       74498$ & $           1$ & $                       \mathrm{Ca}^+.4p$  &$           1$ & $           2$ & $                    1.000$ \\
$                      3d4s.^3D$ & $           2$ & $           3$ & $           4$ & $           0$ & $       20335$ & $       62956$ & $           1$ & $                       \mathrm{Ca}^+.3d$  &$           2$ & $           2$ & $                    1.000$ \\
$                      4s3d.^3D$ & $           2$ & $           3$ & $           3$ & $           2$ & $       20335$ & $       49305$ & $           1$ & $                       \mathrm{Ca}^+.4s$  &$           0$ & $           2$ & $                    1.000$ \\
$                      3d4s.^1D$ & $           2$ & $           1$ & $           4$ & $           0$ & $       21849$ & $       62956$ & $           1$ & $                       \mathrm{Ca}^+.3d$  &$           2$ & $           2$ & $                    1.000$ \\
$                      4s3d.^1D$ & $           2$ & $           1$ & $           3$ & $           2$ & $       21849$ & $       49305$ & $           1$ & $                       \mathrm{Ca}^+.4s$  &$           0$ & $           2$ & $                    1.000$ \\
$                    4s4p.^1P^o$ & $           1$ & $           1$ & $           4$ & $           1$ & $       23652$ & $       49305$ & $           1$ & $                       \mathrm{Ca}^+.4s$  &$           0$ & $           2$ & $                    1.000$ \\
$                    4p4s.^1P^o$ & $           1$ & $           1$ & $           4$ & $           0$ & $       23652$ & $       74498$ & $           1$ & $                       \mathrm{Ca}^+.4p$  &$           1$ & $           2$ & $                    1.000$ \\
$                      4s5s.^3S$ & $           0$ & $           3$ & $           5$ & $           0$ & $       31539$ & $       49305$ & $           1$ & $                       \mathrm{Ca}^+.4s$  &$           0$ & $           2$ & $                    1.000$ \\
$                      4s5s.^1S$ & $           0$ & $           1$ & $           5$ & $           0$ & $       33317$ & $       49305$ & $           1$ & $                       \mathrm{Ca}^+.4s$  &$           0$ & $           2$ & $                    1.000$ \\
$                    3d4p.^3F^o$ & $           3$ & $           3$ & $           4$ & $           1$ & $       35730$ & $       62956$ & $           1$ & $                       \mathrm{Ca}^+.3d$  &$           2$ & $           2$ & $                    1.000$ \\
$                    4p3d.^3F^o$ & $           3$ & $           3$ & $           3$ & $           2$ & $       35730$ & $       74498$ & $           1$ & $                       \mathrm{Ca}^+.4p$  &$           1$ & $           2$ & $                    1.000$ \\
$                    3d4p.^1D^o$ & $           2$ & $           1$ & $           4$ & $           1$ & $       35835$ & $       62956$ & $           1$ & $                       \mathrm{Ca}^+.3d$  &$           2$ & $           2$ & $                    1.000$ \\
$                    4p3d.^1D^o$ & $           2$ & $           1$ & $           3$ & $           2$ & $       35835$ & $       74498$ & $           1$ & $                       \mathrm{Ca}^+.4p$  &$           1$ & $           2$ & $                    1.000$ \\
$                    4s5p.^3P^o$ & $           1$ & $           3$ & $           5$ & $           1$ & $       36547$ & $       49305$ & $           1$ & $                       \mathrm{Ca}^+.4s$  &$           0$ & $           2$ & $                    1.000$ \\
$                    4s5p.^1P^o$ & $           1$ & $           1$ & $           5$ & $           1$ & $       36731$ & $       49305$ & $           1$ & $                       \mathrm{Ca}^+.4s$  &$           0$ & $           2$ & $                    1.000$ \\
$                      4s4d.^1D$ & $           2$ & $           1$ & $           4$ & $           2$ & $       37298$ & $       49305$ & $           1$ & $                       \mathrm{Ca}^+.4s$  &$           0$ & $           2$ & $                    1.000$ \\
$                      4s4d.^3D$ & $           2$ & $           3$ & $           4$ & $           2$ & $       37748$ & $       49305$ & $           1$ & $                       \mathrm{Ca}^+.4s$  &$           0$ & $           2$ & $                    1.000$ \\
$                    3d4p.^3D^o$ & $           2$ & $           3$ & $           4$ & $           1$ & $       38192$ & $       62956$ & $           1$ & $                       \mathrm{Ca}^+.3d$  &$           2$ & $           2$ & $                    1.000$ \\
$                    4p3d.^3D^o$ & $           2$ & $           3$ & $           3$ & $           2$ & $       38192$ & $       74498$ & $           1$ & $                       \mathrm{Ca}^+.4p$  &$           1$ & $           2$ & $                    1.000$ \\
$                      4p^2.^3P$ & $           1$ & $           3$ & $           4$ & $           1$ & $       38417$ & $       74498$ & $           2$ & $                       \mathrm{Ca}^+.4p$  &$           1$ & $           2$ & $                    1.000$ \\
$                    3d4p.^3P^o$ & $           1$ & $           3$ & $           4$ & $           1$ & $       39333$ & $       62956$ & $           1$ & $                       \mathrm{Ca}^+.3d$  &$           2$ & $           2$ & $                    1.000$ \\
$                    4p3d.^3P^o$ & $           1$ & $           3$ & $           3$ & $           2$ & $       39333$ & $       74498$ & $           1$ & $                       \mathrm{Ca}^+.4p$  &$           1$ & $           2$ & $                    1.000$ \\
$                      4s6s.^3S$ & $           0$ & $           3$ & $           6$ & $           0$ & $       40474$ & $       49305$ & $           1$ & $                       \mathrm{Ca}^+.4s$  &$           0$ & $           2$ & $                    1.000$ \\
$\mathrm{Ca}^+(4s)+\mathrm{H}^-$ & $           0$ & $           2$ & $           0$ & $           0$ & $           0$ & $           0$ & $           0$ & $                       \mathrm{Ca}^+.4s$  &$           0$ & $           2$ & $                    0.000$ \\
$\mathrm{Ca}^+(3d)+\mathrm{H}^-$ & $           2$ & $           2$ & $           0$ & $           0$ & $       13650$ & $           0$ & $           0$ & $                       \mathrm{Ca}^+.3d$  &$           2$ & $           2$ & $                    0.000$ \\
$\mathrm{Ca}^+(4p)+\mathrm{H}^-$ & $           1$ & $           2$ & $           0$ & $           0$ & $       25192$ & $           0$ & $           0$ & $                       \mathrm{Ca}^+.4p$  &$           1$ & $           2$ & $                    0.000$ \\
\end{tabular}
\end{ruledtabular}
\end{table*}

\begin{table*}
\caption{\label{tab:ca_syms} Possible symmetries for Ca+H molecular states arising from various asymptotic atomic states, and the total statistical weights.  The symmetries leading to covalent-ionic interactions among the considered states, and thus which need to be calculated, are shown at the bottom along with their statistical weights. In the case of covalent configurations, the hydrogen atom ground state, H($1s$), is implied and for clarity not written.}
\begin{ruledtabular}
\begin{tabular}{rlcl}
Label & Configuration & $g_{total}$& Terms\\ \hline
$  1$ & $                                4s^2.^1S$ &    2& $   ^{2}\Sigma^+$ \\
$  2$ & $                              4s4p.^3P^o$ &   18& $   ^{2}\Sigma^+,\        ^{2}\Pi,\   ^{4}\Sigma^+,\        ^{4}\Pi$ \\
$  3$ & $                                3d4s.^3D$ &   30& $   ^{2}\Sigma^+,\        ^{2}\Pi,\     ^{2}\Delta,\   ^{4}\Sigma^+,\        ^{4}\Pi,\     ^{4}\Delta$ \\
$  4$ & $                                3d4s.^1D$ &   10& $   ^{2}\Sigma^+,\        ^{2}\Pi,\     ^{2}\Delta$ \\
$  5$ & $                              4s4p.^1P^o$ &    6& $   ^{2}\Sigma^+,\        ^{2}\Pi$ \\
$  6$ & $                                4s5s.^3S$ &    6& $   ^{2}\Sigma^+,\   ^{4}\Sigma^+$ \\
$  7$ & $                                4s5s.^1S$ &    2& $   ^{2}\Sigma^+$ \\
$  8$ & $                              3d4p.^3F^o$ &   42& $   ^{2}\Sigma^+,\        ^{2}\Pi,\     ^{2}\Delta,\       ^{2}\Phi,\   ^{4}\Sigma^+,\        ^{4}\Pi,\     ^{4}\Delta,\       ^{4}\Phi$ \\
$  9$ & $                              3d4p.^1D^o$ &   10& $   ^{2}\Sigma^-,\        ^{2}\Pi,\     ^{2}\Delta$ \\
$ 10$ & $                              4s5p.^3P^o$ &   18& $   ^{2}\Sigma^+,\        ^{2}\Pi,\   ^{4}\Sigma^+,\        ^{4}\Pi$ \\
$ 11$ & $                              4s5p.^1P^o$ &    6& $   ^{2}\Sigma^+,\        ^{2}\Pi$ \\
$ 12$ & $                                4s4d.^1D$ &   10& $   ^{2}\Sigma^+,\        ^{2}\Pi,\     ^{2}\Delta$ \\
$ 13$ & $                                4s4d.^3D$ &   30& $   ^{2}\Sigma^+,\        ^{2}\Pi,\     ^{2}\Delta,\   ^{4}\Sigma^+,\        ^{4}\Pi,\     ^{4}\Delta$ \\
$ 14$ & $                              3d4p.^3D^o$ &   30& $   ^{2}\Sigma^-,\        ^{2}\Pi,\     ^{2}\Delta,\   ^{4}\Sigma^-,\        ^{4}\Pi,\     ^{4}\Delta$ \\
$ 15$ & $                                4p^2.^3P$ &   18& $   ^{2}\Sigma^-,\        ^{2}\Pi,\   ^{4}\Sigma^-,\        ^{4}\Pi$ \\
$ 16$ & $                              3d4p.^3P^o$ &   18& $   ^{2}\Sigma^+,\        ^{2}\Pi,\   ^{4}\Sigma^+,\        ^{4}\Pi$ \\
$ 17$ & $                                4s6s.^3S$ &    6& $   ^{2}\Sigma^+,\   ^{4}\Sigma^+$ \\
$ 18$ & $          \mathrm{Ca}^+(4s)+\mathrm{H}^-$ &    2& $   ^{2}\Sigma^+$ \\
$ 19$ & $          \mathrm{Ca}^+(3d)+\mathrm{H}^-$ &   10& $   ^{2}\Sigma^+,\        ^{2}\Pi,\     ^{2}\Delta$ \\
$ 20$ & $          \mathrm{Ca}^+(4p)+\mathrm{H}^-$ &    6& $   ^{2}\Sigma^+,\        ^{2}\Pi$ \\
&&&\\ 
     \multicolumn{3}{r}{Number of symmetries to calculate :   3} & $    ^{2}\Sigma^+,\        ^{2}\Pi,\     ^{2}\Delta$ \\
 \multicolumn{3}{r}{$g:$} & $              \makebox[7mm][r]{2,}\             \makebox[5mm][r]{4,}\              \makebox[5mm][r]{4}$ \\
\end{tabular}
\end{ruledtabular}
\end{table*}

\begin{figure*}
\centering
\includegraphics[width=1.0\textwidth]{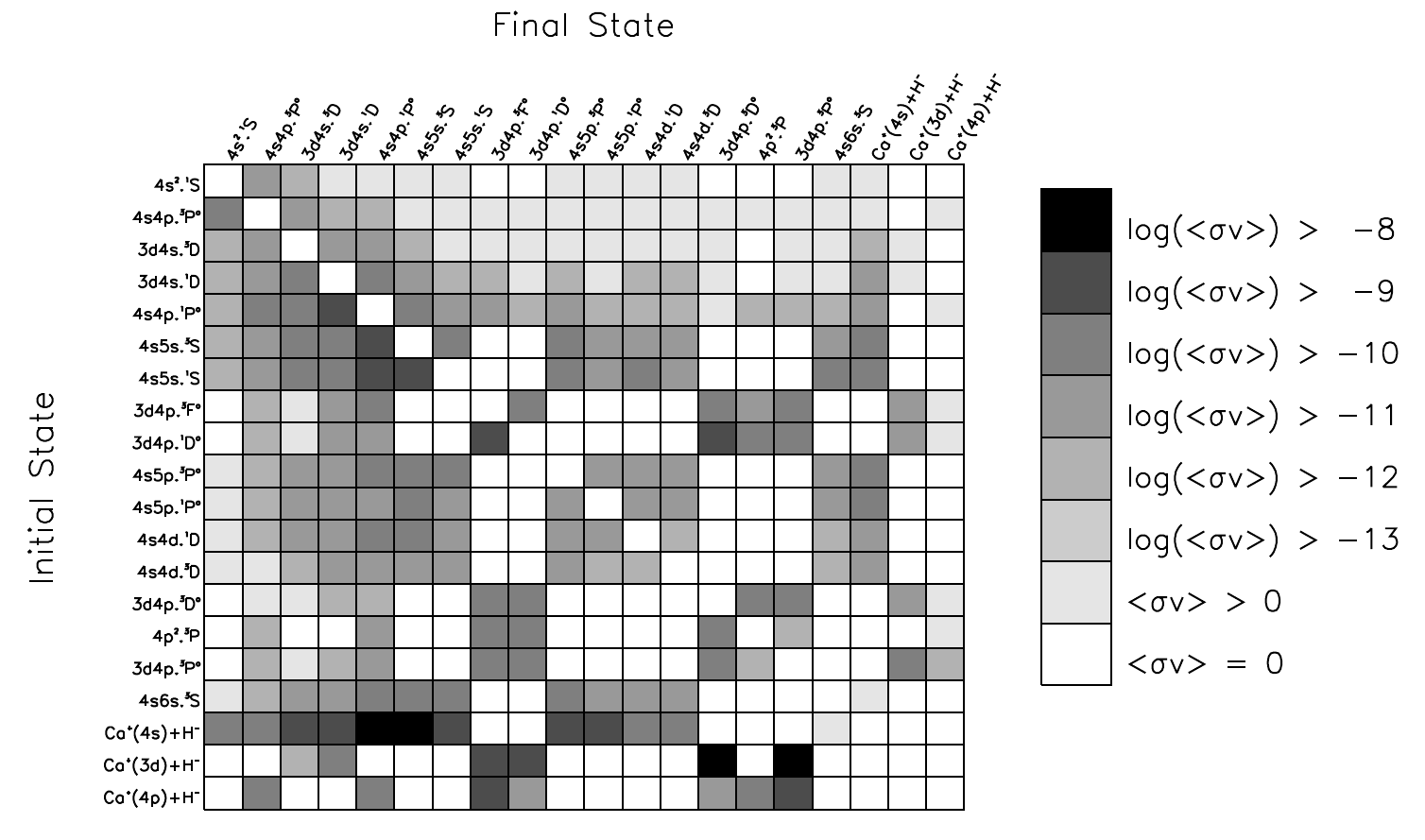}
\caption{Graphical representation of the rate coefficient matrix  $\la \sigma v \ra$ (in cm$^3$~s$^{-1}$) for inelastic Ca + H and Ca$^+$ + H$^-$ collisions at temperature $T = 6000$~K.  Results are from the LCAO asymptotic model.  The logarithms in the legend are to base 10. }\label{fig:Ca_rates_grid}
\end{figure*}

\begin{figure}
\centering
\includegraphics[width=0.5\textwidth]{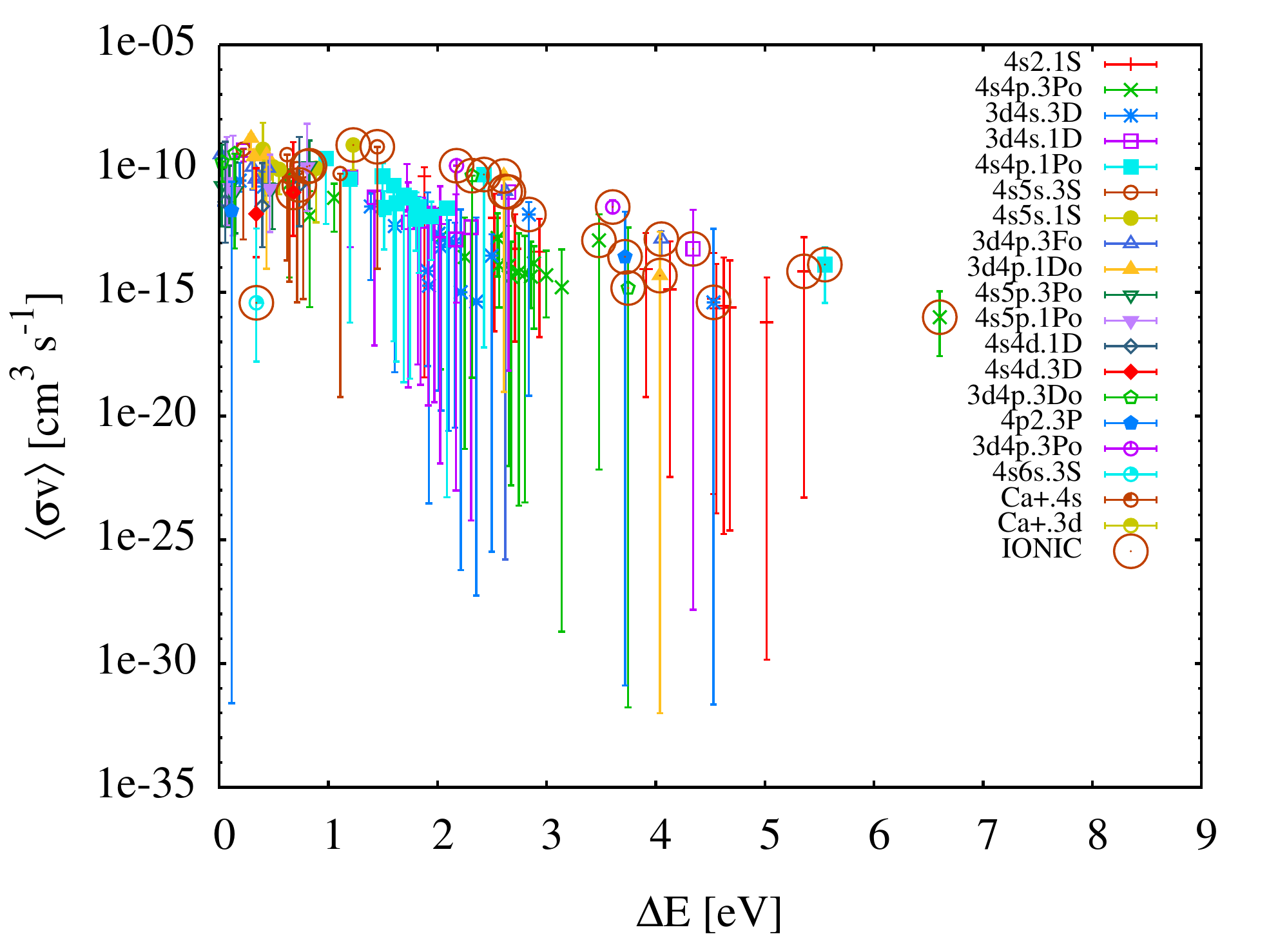}
\caption{The rate coefficients $\la \sigma v \ra$ for Ca+H collision processes at 6000~K, plotted against the asymptotic energy difference between initial and final molecular states, $\Delta E$.  The data are shown for endothermic processes. i.e. excitation and ion-pair production.  The legend labels the initial state of the transition, and processes leading to a final ionic state (ion-pair production) are circled.  The points show the results of the LCAO asymptotic model, with the bars showing the fluctuations.}\label{fig:Ca_rates}
\end{figure}

The rate coefficients are calculated for temperatures in the range 1000--20000~K, with steps of 1000~K, for the various models.  The LCAO model results, as well as the minimum and maximum values from alternate models, the fluctuations, are published electronically as Supplemental Material \footnote{See Supplemental Material at [URL will be inserted by publisher] for tables of rate coefficients.}.  Example results at 6000~K, corresponding to a typical spectrum forming region in a solar-type star, are shown in Figs.~\ref{fig:Ca_rates_grid} and \ref{fig:Ca_rates}.  The first thing to notice from the figures, is that, as has been found in Li+H, Na+H, and Mg+H, that the largest rate coefficients for processes involving the ground state core $\mathrm{Ca}^+(4s)$ correspond to charge transfer processes and excitation between neighbouring states involving moderately excited states near the first excited S-state, here in particular $4s4p \, ^1$P$^o$, $4s5s \, ^3$S, and $4s5s \, ^1$S.  The reasons that the first excited $S$-state generally provides the largest rate coefficients are that the first excited states lead to ionic crossings at intermediate internuclear distances where the transition probability becomes optimal, and that S-states lead to the largest statistical weights for the initial channels \cite{Barklem2012}.  We also note that the charge transfer processes involving the $\mathrm{Ca}^+(3d)$ core and the $3d4p \, ^3$D$^o$, and $3d4p \, ^3$P$^o$ states are also significant.  Fig.~\ref{fig:Ca_rates} also demonstrates that the processes with the largest rate coefficients, as for Li+H, Na+H, and Mg+H, tend to have the smallest fluctuations, often around one order of magnitude.  Other processes can have very large fluctuations of many orders of magnitude, but the maximum values are not large enough that the process is likely to be important in applications.

\section{Conclusions}

An asymptotic two-electron LCAO method for the treatment of ionic-covalent interactions has been developed, based on earlier work by GAH and Anstee.  The method presented here makes several improvements, including the extension to complex atoms and the use of fully anti-symmetrised wavefunctions.  When coupled with standard multi-channel Landau-Zener formulae for treatment of the dynamics, the method allows rates for excitation and charge transfer processes in low-energy hydrogen atom collisions with neutral atoms to be estimated.  The results for Li+H, Na+H, and Mg+H compare well with the existing detailed full-quantum studies.  The LCAO method clearly outperforms alternate models based on semi-empirical or Landau-Herring estimates of couplings.  The alternate model calculations are useful, however, to provide a measure of the sensitivity of the calculations to the couplings, and are thus used to calculate ``fluctuations'', which are analogous to an estimate of the uncertainty.  In this paper we have chosen to focus on comparison with the full-quantum calculations for Li+H, Na+H, and Mg+H collisions, as they provide the most robust test of the method.  In the near future we plan to calculate data for Al+H and Si+H, and thus to compare with semi-empirical model calculations \cite{Belyaev2013,Belyaev2013a,Belyaev2014}.  The calculation of adiabatic radial couplings using the method was also demonstrated, and may be useful for hybrid calculations combining model and full-quantum approaches. 

The method has been used to calculate data for the astrophysically important case of Ca+H collisions.  As found in Li+H, Na+H, and Mg+H, the largest rate coefficients for processes involving the ground state core correspond to charge transfer processes and excitation between neighbouring states involving moderately excited states near the first excited S-state.  In Ca+H, charge transfer processes involving an excited core configuration and more excited states have also been found to be important.  Processes with the largest rate coefficients tend to have the smallest fluctuations, often around one order of magnitude.  Other processes can have very large fluctuations, but the maximum values are not large compared to the rate coefficients for the most efficient processes, and are thus unlikely to be astrophysically important.  This can be checked in astrophysical modelling by allowing rate coefficients to vary according to the fluctuations.  Such calculations should be used to assess if the process and its estimated uncertainty could possibly be important, and thus if more accurate calculations are required.   The described method will be applied in the near future to various other cases of astrophysical interest for which data do not exist, e.g. Fe+H, K+H, Ti+H.  We note that the described method cannot be applied directly to open-shell elements such as oxygen.  The large ionisation potential means that configurations involving excited states of hydrogen must be included.

\begin{acknowledgments}
I thank Andrey Belyaev for many years of fruitful collaboration and enlightening discussions, without which this work would not have been possible.  I thank him also for comments on this manuscript. I thank Stuart Anstee for making available to me his codes, on which this work was built.  This work received financial support from the Royal Swedish Academy of Sciences, the Wenner-Gren Foundation, Göran Gustafssons Stiftelse and the Swedish Research Council.  For much of this work I was a Royal Swedish Academy of Sciences Research Fellow supported by a grant from the Knut and Alice Wallenberg Foundation. I am presently partially supported by the project grant ``The New Milky Way'' from the Knut and Alice Wallenberg Foundation.
\end{acknowledgments}

\appendix

\section{Matrix elements}
\label{app:mat}

\subsection{Two-electron matrix elements}
\label{app:twomat}

Here we derive expressions for the matrix elements of the Hamiltonian, in terms of component matrix elements and overlap matrix elements, which will be derived below.   We need four different matrix elements $H_{11}^\mathrm{NC}$, $H_{1j}^\mathrm{NC}$, $H_{jj}^\mathrm{NC}$ and $H_{jk}^\mathrm{NC}$.   In deriving the matrix elements we make use of the appropriate partitioning of $H$, which leads to simplifications via $H^\mathrm{ion} | \varphi^\mathrm{A}_{1s} \varphi^\mathrm{A}_{LR} \ra = E^\mathrm{ion}| \varphi^\mathrm{A}_{1s} \varphi^\mathrm{A}_{LR} \ra $, where $E^\mathrm{ion}$ is the asymptotic ion-pair energy, and $H^\mathrm{cov} | \varphi^\mathrm{A}_{1s} \varphi^\mathrm{B}_j \ra = E_j| \varphi^\mathrm{A}_{1s} \varphi^\mathrm{B}_j \ra$, where $E_j$ is the asymptotic energy of the covalent state where the hydrogen atom on A is in the $1s$ state and atom X on  B is in state $j$.

Our expressions differ somewhat from those of GAH and \cite{Anstee1992}.  First, in both cases differences occur due to the fact that we have employed a correctly anti-symmetrised ionic wavefunction, eqn~\ref{eq:ionicwf}.  The new expressions are not significantly more difficult to compute than those derived by \cite{Anstee1992}.
Second, as done by \cite{Anstee1992}, we have retained many terms that go asymptotically to zero and consequently that were neglected by GAH.  Once again, we have not found retaining these terms to increase the difficulty of the calculations significantly, and so we choose to retain them.

The required matrix elements of the Hamiltonian are written as follows.  First, the diagonal element involving the ionic state
\begin{widetext}
\bea
H_{11}^\mathrm{NC} & = & \la \Phi_1^\mathrm{NC} | H | \Phi_1^\mathrm{NC} \ra \n 
       & = & \frac{1}{2(1+S_{0L}^2)} \left\{ \la \varphi^\mathrm{A}_{1s} \varphi^\mathrm{A}_{LR} |H| \varphi^\mathrm{A}_{1s} \varphi^\mathrm{A}_{LR} \ra 
             +  \la \varphi^\mathrm{A}_{1s} \varphi^\mathrm{A}_{LR} |H| \varphi^\mathrm{A}_{LR} \varphi^\mathrm{A}_{1s}  \ra  
             +  \la \varphi^\mathrm{A}_{LR} \varphi^\mathrm{A}_{1s} |H| \varphi^\mathrm{A}_{1s} \varphi^\mathrm{A}_{LR}  \ra  
             +  \la \varphi^\mathrm{A}_{LR} \varphi^\mathrm{A}_{1s} |H| \varphi^\mathrm{A}_{LR} \varphi^\mathrm{A}_{1s} \ra  \right\} \n 
       & = &  (E^\mathrm{ion} + \frac{1}{R}) - \frac{1}{(1+S_{0L}^2)} \left\{  \la \varphi^\mathrm{A}_{1s} | \frac{1}{r_\mathrm{B}} | \varphi^\mathrm{A}_{1s} \ra + \la \varphi^\mathrm{A}_{LR} | \frac{1}{r_\mathrm{B}} | \varphi^\mathrm{A}_{LR} \ra + 2 \la \varphi^\mathrm{A}_{1s} | \frac{1}{r_\mathrm{B}} | \varphi^\mathrm{A}_{LR} \ra S_{0L} \right\}.                
\eea
At long range as $r_\mathrm{B} \rightarrow R$ for matrix elements on A and $S_{0L} \rightarrow 0$, we recover the expression of GAH
\beq
H_{11}^\mathrm{NC}  \sim  E^\mathrm{ion} - \frac{1}{R}.
\eeq
Second, the off-diagonal element involving an ionic and a covalent state
\bea
H_{1j}^\mathrm{NC} & = & \la \Phi_1^\mathrm{NC} | H | \Phi_j^\mathrm{NC} \ra \n
       & = & \frac{1}{2\sqrt{(1+S_{0j}^2)(1+S_{0L}^2)}} \left\{ \la \varphi^\mathrm{A}_{1s} \varphi^\mathrm{A}_{LR} |H| \varphi^\mathrm{A}_{1s} \varphi^\mathrm{B}_j \ra 
             +  \la \varphi^\mathrm{A}_{1s} \varphi^\mathrm{A}_{LR} |H| \varphi^\mathrm{B}_j \varphi^\mathrm{A}_{1s}  \ra +  \la \varphi^\mathrm{A}_{LR} \varphi^\mathrm{A}_{1s} |H| \varphi^\mathrm{A}_{1s} \varphi^\mathrm{B}_j  \ra  
             +  \la \varphi^\mathrm{A}_{LR} \varphi^\mathrm{A}_{1s} |H| \varphi^\mathrm{B}_j \varphi^\mathrm{A}_{1s} \ra  \right\} \n 
       & = & (E^\mathrm{ion} + \frac{1}{R}) {S}_{1j}^\mathrm{NC} - \frac{1}{2\sqrt{(1+S_{0j}^2)(1+S_{0L}^2)}} \left\{\la \varphi^\mathrm{A}_{1s} | \frac{1}{r_\mathrm{B}} | \varphi^\mathrm{A}_{1s} \ra S_{jL} + \la \varphi^\mathrm{A}_{LR} | \frac{1}{r_\mathrm{B}} | \varphi^\mathrm{B}_j \ra + \la \varphi^\mathrm{A}_{1s} | \frac{1}{r_\mathrm{B}} | \varphi^\mathrm{B}_j \ra S_{0L} \right. \n && \left. + \la \varphi^\mathrm{A}_{LR} | \frac{1}{r_\mathrm{B}} | \varphi^\mathrm{A}_{1s} \ra S_{0j} \right\}. \label{eq:ionHcov}
\eea
\end{widetext}
Third, the diagonal element involving a covalent state
\bea
H_{jj}^\mathrm{NC} & = & \la \Phi_j^\mathrm{NC} | H | \Phi_j^\mathrm{NC} \ra \n
       & = &  E_j + \la \textrm{covalent interaction} \ra \n
       & \approx & E_j, 
\eea
the approximation being valid at long range.  Finally, the off-diagonal matrix elements involving two covalent states
\bea
H_{jk}^\mathrm{NC} & = & \la \Phi_j^\mathrm{NC} | H | \Phi_k^\mathrm{NC} \ra \n
       & \approx & \frac{1}{2}(E_j + E_k) {S}^\mathrm{NC}_{jk}, \label{eq:covHcov}
\eea
the approximation being valid at long range.  For an orthonormal set of functions ${S}^\mathrm{NC}_{jk} = 0$, and so these matrix elements are assumed zero.

We choose the zero point energy to be the asymptotic energy corresponding to both atoms in their ground states, and thus
\beq
E_j = E_0 + E_j^X =  E_j^X,
\eeq
where $E_0$ is the ground state energy of the hydrogen atom on A, and $E_j^X$ is the energy of atomic state of atom X corresponding to $\varphi^\mathrm{B}_j$ with respect to its ground state.  Additionally, the ion-pair limit can be written
\beq
E^\mathrm{ion} = E^{X^+} - E^{H^-}
\eeq
where $E^{X^+}$ is the series limit for the considered atom X (i.e. for a given core configuration X$^+(nl)$) relative to the ground state of X, and is equivalent to $E_{lim}$ discussed in the main text.  In the case that the considered core configuration corresponds to that of the ground state, this corresponds to the ionisation energy.  $E^{H^-}$ is the electron affinity of hydrogen.

The two-electron overlap integrals $S_{1j}^\mathrm{NC}$ are needed both in the calculation of the Hamiltonian above (eqn.~\ref{eq:ionHcov}) and for the calculation of the overlap matrix needed for eqn~\ref{eq:genmat}.  These can be written in terms of one-electron overlap integrals, which will be given in \ref{app:onemat}.  The overlap between the ionic and covalent states is given by
\bea
S^\mathrm{NC}_{1k} & = & \la \Phi_1^\mathrm{NC} | \Phi_k^\mathrm{NC} \ra \n
& = & \frac{S_{kL} + S_{01} S_{0k}}{\sqrt{(1+S_{01}^2)(1+S_{0k}^2)}}.
\eea
Note, the overlap between covalent states is given by 
\bea
S^\mathrm{NC}_{jk} & = & \la \Phi_j^\mathrm{NC} | \Phi_k^\mathrm{NC} \ra \n
& = &  \frac{S_{jk} + S_{0j} S_{0k}}{\sqrt{(1+S_{0j}^2)(1+S_{0k}^2)}}.
\eea
For an orthonormal basis the one-electron overlap $S_{jk}=\delta_{jk}$, and the remaining terms are asymptotically very small.  We therefore assume ${S}^\mathrm{NC}_{jk}\approx\delta_{jk}$.

\subsection{One-electron matrix elements}
\label{app:onemat}

The two-electron Hamiltonian matrix elements and overlap matrix elements require the calculation of various one electron matrix elements and overlaps.
Four different one-electron overlaps are required, $S_{0L}$, $S_{0j}$, $S_{jL}$ and $S_{jk}$.  There are four different one-electron overlaps potentially required
\begin{enumerate}
\item
$S_{0L} = \br \varphi^\mathrm{A}_{1s}|\varphi^\mathrm{A}_{LR}\ke$
\item
$S_{0j} = \br \varphi^\mathrm{A}_{1s}|\varphi^\mathrm{B}_j\ke$
\item
$S_{jL} = \br \varphi^\mathrm{B}_k|\varphi^\mathrm{A}_{LR}\ke$
\item
$S_{jk} = \br \varphi^\mathrm{B}_j| \varphi^\mathrm{B}_k\ke$
\end{enumerate}
The first overlap involves two orbitals on atom A, and the last overlap two orbitals on B, and thus are independent of $R$.   In \cite{Anstee1992}, Anstee derived the analytic expression for overlap 1 depending on the parameters $r_0$ and $\gamma$ in the function $\varphi^\mathrm{A}_{LR}$, which for our choices gives a value of $0.859$.   The remaining three overlap integrals involve $\varphi^\mathrm{B}_j$.   The two-electron overlap between covalent states ${S}_{jk}^\mathrm{NC}$ is assumed asymptotically to go to zero, and thus ${S}_{jk}$ is not required, though it would be zero anyway if set of functions $\varphi^\mathrm{B}_j$ is orthogonal; but this is not necessarily guaranteed for our method.  The remaining two, $S_{0j}$ and $S_{jL}$, can be integrated analytically as far as possible, but require a final numerical integration over the radial part of the wavefunction $P_{nl}(r_\mathrm{B})$.  Note, these overlaps are only non-zero if $m=0$.  For the $m=0$ case, angular parts of the wavefunction depend on the $l$ quantum number, and different expressions are derived for different $l$.  In \cite{Anstee1992}, Anstee derived expressions for $l=0$ and $l=1$.  Test calculations adopting $l=0$ expressions for all values of $l$ indicate, that accounting for $l$ correctly is not greatly important.  We adopt the expression for $l=1$ for all cases where $l\ne 0$. 

There are five component matrix elements required are:
\begin{enumerate}
\item
$\br \varphi^\mathrm{A}_{1s}|\dfrac{1}{r_\mathrm{B}}|\varphi^\mathrm{A}_{1s}\ke $
\item
$\br \varphi^\mathrm{A}_{LR}|\dfrac{1}{r_\mathrm{B}}|\varphi^\mathrm{A}_{LR}\ke$
\item
$\br \varphi^\mathrm{A}_{1s}|\dfrac{1}{r_\mathrm{B}}|\varphi^\mathrm{A}_{LR}\ke$
\item
$\br \varphi^\mathrm{A}_{LR}|\dfrac{1}{r_\mathrm{B}}|\varphi^\mathrm{B}_j\ke$
\item
$\br \varphi^\mathrm{A}_{1s}|\dfrac{1}{r_\mathrm{B}}| \varphi^\mathrm{B}_j\ke$
\end{enumerate}
For matrix elements 1-3, analytic expressions have been derived by \cite{Anstee1992}.  Elements 4-5 are calculated numerically similar to the corresponding overlaps $S_{jL}$ and $S_{0j}$, with an extra $1/r_\mathrm{B}$ factor in the numerical integral. 

\section{Radial couplings}
\label{app:radcoup}

The method for calculating the non-adiabatic radial couplings from the two-elecron LCAO model is now described.  It should be emphasised that the described approach is approximate, and in particular the choice of coordinate system for the electrons can play an important role in the correct calculation and use of radial couplings \cite{Allan1983,belyaev_dependence_2002}.  In particular, to avoid problems Jacobi coordinates should be preferred.  However, such effects are expected to be smaller than those due to other approximations in the model.  In any case, the approximate radial coupling in this model may be calculated directly in terms of numerical derivatives of existing matrices.  Our method is similar to that of \cite{boutalib_abinitio_1992}, who used two methods to calculate radial couplings: direct differentiation of the coefficient matrix $\mathbf{c}$, and the Hellmann-Feynman relation \cite{hellmann_einfuhrung_1944,feynman_forces_1939}.  We note that the expressions derived by \cite{boutalib_abinitio_1992} are only applicable in the case of a strict diabatisation: the diabatic states are orthonormal (i.e. the overlap matrix is the identity matrix, $S=I$) and that the radial couplings in the diabatic representation are zero (i.e. $\la \Phi_i | \frac{\partial}{ \partial R} | \Phi_j \ra = 0$).  These conditions are not fulfilled in the asymptotic model used here, and so we derive appropriate expressions for the direct differentiation case in the asymptotic model.

We write the adiabatic radial couplings in terms of the diabatic radial couplings via the $c$ coefficients
\bea
\la \Psi_i | \frac{\partial}{\partial R} | \Psi_j \ra & = & \sum_k \sum_l c_{ki}^* \la \Phi_k | \frac{\partial}{\partial R} | \Phi_l \ra c_{lj} \n 
& = & \sum_k \sum_l c_{ki} \la \Phi_k | \frac{\partial}{\partial R} | \Phi_l \ra c_{lj}.
\eea
Using the product rule
\bea
\la \Psi_i | \frac{\partial}{\partial R} | \Psi_j \ra & = & \sum_k \sum_l c_{ki} \la \Phi_k | \left[ | \Phi_l \ra \frac{\partial c_{lj}}{\partial R}  +  \frac{\partial | \Phi_l \ra}{\partial R} c_{lj} \right] \n 
& = & \sum_k \sum_l c_{ki} \left[ {S}_{kl} \frac{\partial c_{lj}}{\partial R}  +  c_{lj} \la \Phi_k | \frac{\partial }{\partial R}| \Phi_l \ra  \right]. \n &&\label{eq:rc1}
\eea
In the limit that $\mathbf{S}=I$ and that $\la \Phi_i | \partial / \partial R | \Phi_j \ra = 0$, i.e. strict diabatisation, we recover the expression of \cite{boutalib_abinitio_1992}
\beq
\la \Psi_i | \frac{\partial}{\partial R} | \Psi_j \ra  = \sum_k c_{ki} \frac{\partial c_{kj}}{\partial R}, 
\eeq
or in matrix notation
\beq
\la \Psi_i | \frac{\partial}{\partial R} | \Psi_j \ra  = \left\langle \mathbf{c}^\dagger \frac{\partial \mathbf{c}}{\partial R} \right\rangle_{ij} .
\eeq
These conditions are not applicable for our basis set, in particular there are significant overlaps and couplings between ionic and covalent states, and thus we must calculate eqn~\ref{eq:rc1}.  It is helpful to consider the product rule
\bea
\frac{\partial}{\partial R} \la \Phi_k | \Phi_l \ra & = & \left(\frac{\partial }{\partial R} \la \Phi_k | \right)  | \Phi_l \ra + \la \Phi_k | \frac{\partial }{\partial R} | \Phi_l \ra \n 
& = & \la \Phi_l |\frac{\partial }{\partial R}  | \Phi_k \ra + \la \Phi_k | \frac{\partial }{\partial R}| \Phi_l \ra, \label{eqn:drc1}
\eea
which thus relates the couplings to the derivative of the overlap matrix elements.  In the case that the basis is the adiabatic one, the overlap matrix on the left-hand side becomes the identity matrix and one recovers the expected properties of the radial coupling matrix in the adiabatic basis:
\beq
\la \Psi_i | \frac{\partial}{\partial R} | \Psi_i \ra = 0, \label{eqn:arc1}
\eeq
and 
\beq
\la \Psi_i | \frac{\partial}{\partial R} | \Psi_j \ra = - \la \Psi_j | \frac{\partial}{\partial R} | \Psi_i \ra . \label{eqn:arc2}
\eeq
In the diabatic basis used here the overlap matrix has non-zero off-diagonal elements.  In the case that one of the states is the ionic state $| \Phi_1 \ra$, and taking the origin of electronic coordinates on the hydrogen atom on A, since this state has no dependence on R, we obtain
\beq
 \la \Phi_k | \frac{\partial }{\partial R} | \Phi_1 \ra  =  0, \;
 \la \Phi_1 | \frac{\partial }{\partial R} | \Phi_k \ra  =  \frac{\partial S_{1k}}{\partial R}.
\eeq
Taking instead the origin of electronic coordinates on B, we obtain the asymptotic results
\beq
 \la \Phi_k | \frac{\partial }{\partial R} | \Phi_1 \ra  \approx  \frac{\partial S_{1k}}{\partial R}, \;
 \la \Phi_1 | \frac{\partial }{\partial R} | \Phi_k \ra  \approx  0.
\eeq
In the case where both states are covalent, eqn \ref{eqn:drc1} suggests the diabatic couplings are of order
\beq
 \la \Phi_k | \frac{\partial }{\partial R} | \Phi_l \ra \sim \la \Phi_l | \frac{\partial }{\partial R} | \Phi_k \ra \sim \frac{1}{2} \frac{\partial {S}_{lk}}{\partial R},  \label{eqn:rcapprox}
\eeq
and will be very small at large $R$, noting ${S}_{lk}\approx 0$ asymptotically.  The coefficients for such terms, $c_{ik} c_{lj}$, will also be very small asymptotically, and thus these terms may be neglected.  In this approximation, taking the case of origin of electronic coordinates on B, we obtain
\beq
\la \Psi_i | \frac{\partial}{\partial R} | \Psi_j \ra 
\approx \sum_k \sum_l c_{ik} {S}_{kl} \frac{\partial c_{lj}}{\partial R}  +  \sum_k c_{ik} c_{1j}    \frac{\partial {S}_{1k}}{\partial R} . \label{eq:rcf}
\eeq
Calculations show the approximation in eqns \ref{eqn:rcapprox} and \ref{eq:rcf} works well in the sense that it produces a radial coupling matrix in the adiabatic basis with the expected anti-symmetric properties, eqns \ref{eqn:arc1} and \ref{eqn:arc2}, to quite high precision at the internuclear distances of interest.

\bibliography{../../MyLibrary}

\end{document}